# Advances in 2D Based Field Effect Transistors as Biosensing Platforms: From Principle to Biomedical Applications

Foad Ghasemi[1]*, Abdollah Salimi[2,3]*

[1]Nanoscale Physics Device Lab (NPDL), Department of Physics, University of Kurdistan, 66177-15175, Sanandaj, Iran.

[2]Department of Chemistry, University of Kurdistan, 66177-15175, Sanandaj-Iran.

[3]Research Center for Nanotechnology, University of Kurdistan, 66177-15175, Sanandaj-Iran.

*Corresponding Authors

f.ghasemi@uok.ac.ir, absalimi@uok.ac.ir,

**Abstract**:

Two-dimensional (2D) materials have been used extensively in various fields due to their unique physical and chemical properties. Among their diverse applications, field-effect transistor biosensors (bio-FETs) promise a brilliant prospect in the fabrication of biodevices for diagnostics especially point of care (PoC) based biomedical testing. The introduction of 2D nanomaterials as a sensing platform not only promotes their bioassay process but also provides label-free, low-cost, real-time, highly sensitive, and selective detection of different biomolecules. 2D-based bio-FETs are highly desirable for biosensing applications due to large effective surface area, environment-dependent conductivity, being easily functionalized, biocompatible and simple fabrication processes. This review focuses comprehensively on 2D bio-FETs and more on both dry and wet states. In this regard, these bio-FETs are classified into four categories including Apta, DNA (Geno), Immuno, and enzymatic biosensors. The configuration, detection mechanism and electrical signal processing of each category are discussed in response to different biotargets. Finally, their promising roles as PoC testing devices for modern medicine are discussed in biomedical (real) samples.

## 1. Introduction

Human health is a critical key to the continuity of life, which attempts to improve it resulting in an increase in life quality [1]. Several factors can endanger human health, the most important of them are diseases [2]. As a result, the diagnostic and treatment of diseases play essential roles in protecting human health[3]. Until now, various methods have been introduced to diagnose diseases such a: chemical testing [4], medical imaging [5], spectroscopic techniques [6], electrodiagnostic medicine [7], DNA testing [8], visual examination [9], and physical examination [10]. Observational or physical methods usually come with less accuracy and in special cases, more advanced imaging or spectroscopic techniques are required [11]. However, these methods are time-consuming, labor intensive, high cost, and in some cases associated with low sensitivities [12, 13]. Therefore, significant efforts are ongoing to introduce effective, efficient and inexpensive medical methods.

In the last decade, biosensors demonstrate a superior candidate for disease diagnostic purposes. A biosensor is defined as an analytical device consisting of detector and biomolecule recognition components [14]. The latter part is served to interact with the target analyst and the detector converts the biological response into the desired signals [14]. Based on detection mechanism, biosensors categorize into electrochemical, fluorescent, colorimetric, field effect transistor (FET), surface plasmon resonance (SPR), and piezoelectric biosensors [15-18]. They can recognize various biomolecules including enzymes, carbohydrates (glucose, uric acid, lactic acid), lipids (Cholesterol), proteins, antibodies, viruses, bacteria, and nucleic acids (RNA, DNA) [19-24]. However, practical applications of some biosensors are limited due to lack of selectivity, low stability, and being as expensive devices [25-28].

Among them, FET biosensors have attracted considerable attention because they not only promise sensitive, selective, label-free, early, and fast detection of biomolecules, but also are compatible with silicon industry [29-32]. A FET biosensor is composed of a semiconductor channel and a gate electrode separated by a dielectric from the channel[33, 34]. When biomolecules are introduced into the bio-FET, their corresponding interactions change carrier distributions of the channel resulting in a change of the electrical conductivity [35]. FETs are highly sensitive to their surroundings, however, they offer a low selectivity. In order to increase the selectivity of the FET biosensor, biomolecules called probes are immobilized on their surface through appropriate linkers. Depending on the probe type, the FET biosensors are classified into four categories of aptamer, DNA, enzymatic, and immune FET biosensors (**Figure 1**).

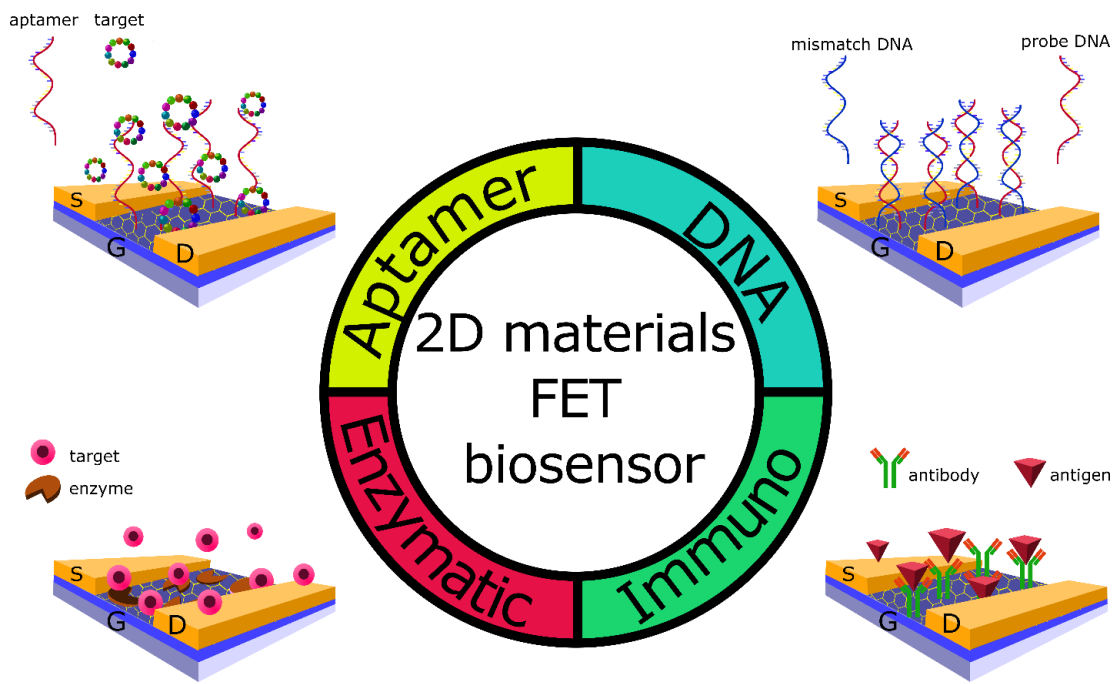

**Figure 1.** 2D based aptamer, DNA, immuno, and enzymatic FET biosensors.

To date, a variety of semiconductor materials have been used as a channel in the FET biosensors including metal oxides, carbon nanotubes, polymer-based compounds, inorganic materials, and various composites [35-39]. In 2004, a carbon monolayer known as graphene was discovered and has brought a great revolution in the field of material science [40]. Distinct properties of the one-atom-thick graphene attract increasing attention among research communities. The large surface area and high electrical conductivity of graphene promote it as a promising candidate for sensing applications [41, 42]. However, graphene is not alone and today, the 2D material family consists of numerous structures with unique properties [43-49]. Large detection surface area, remarkable carrier field-effect mobility, high sensitivity, acceptable biocompatibility, and easy surface functionalization of the 2D based FETs make them a superior candidate for biosensing applications [50-54].

This review focuses on 2D based bio-FETs and their related configurations. The basis of operation of the bio-FETs, target identifications, and signal detection are thoroughly discussed. Many of these biosensors perform detection in aqueous solutions where the Debye length, the isoelectric point and pH play vital roles in affecting the final performance of the bio-FETs. Therefore, these essential concepts are introduced and their impact on the identification of biotargets is discussed. Two basic mechanisms of biotarget detection named doping and gating charge transfer are explained in this class of biosensors. Then, different 2D based bio-FETs are presented and their merits are accurately compared with each other. One of the biggest challenges facing these biosensors is the ability to detect targets in biomedical samples. Therefore, in the final part of this review, the clinical performance of the 2D based bio-FETs are discussed where promises a great development of the PoC devices.

1. **2D bio-FET configurations**

Most 2D bio-FETs include three main parts: 1) Source/drain contacts 2) 2D channel composed of the sensitive biological material and 3) gate electrode (solution or back gates). Source/drain contacts can be metallic or even 2D conductive materials such as graphene. These electrodes are responsible for supplying electrical charge within the channel. The 2D material serves as a channel of the bio-FET and is responsible for the recognition of target biomolecules. For this purpose, the 2D layer is first functionalized with appropriate linkers in order to immobilize probes that are responsible for binding with the target biomolecules[34]. The detection mechanism can be based on direct charges transfer, induction of an electrostatic voltage, redox reactions, or any other interactions between biotargets and 2D layers that lead to a shift in the electrical conductivity of the FETs [55-58]. The third part of the bio-FETs, gate electrode can be a back- or a solution- gate called dry- or wet- states, respectively[59]. In dry-state operation (**Figure 2a**), an oxide layer acts as a dielectric, which is embedded beneath or above of the channel [59]. In wet-state operation (**Figure 2b**), a buffer solution acts as a dielectric and gate voltages are commonly applied through metallic electrodes such as Ag/AgCl, Ag, Pt wires, and so on [59-61].

The main detection mechanisms in 2D bio-FETs are based on two principles: doping effect[62-65] and electrostatic gating effect[66, 67]. In the doping effect, a direct charge transfer occurs between the probe-target duplex and 2D channel material (**Figure 2c**). As an example, in a pristine graphene-based FET (GFET), the minimum conductivity or called Dirac point ($\sigma_{min}$) is generally observed at zero gate voltage ($V_g=0$). However, it usually involves a shift toward positive or negative voltages after being functionalized with linkers or being immobilized with probes. Regardless of this shift, as the probe-target binding occurs, this interaction may donate electrons (called n-doping) or accept electrons (called p-doping) [62-65]. The n-doping biotargets drive the

$\sigma_{min}$ to the negative gate voltages[62, 65], while the p-doping biotargets move the Dirac point to the positive gate voltages (**Figure 2d**)[63, 64]. In contrast, the gating effect appears due to accumulation of charges on the surface of the channel arising from probe-target binding which create a local external voltage drop across the channel[66, 67]. In the case of GFET, positively charged biotargets lead to a shift of $\sigma_{min}$ to the negative gate voltages, while the negatively charged biotargets result in a transfer of $\sigma_{min}$ to the positive gate voltages (**Figure 2e**) [66, 67]. In practice, both mechanisms may be involved in the biotarget sensing process that directly affect the drain current of the bio-FETs.

Among the various configurations of the 2D bio-FETs, electrolyte-gated FETs have attracted more attention thanks to their higher sensing performances compared with the dry-state cases [59]. They could be also integrated with a Polydimethylsiloxane (PDMS) microchannel or well to provide an effective target delivery on the surface of the channel[68]. Due to lower electrical resistances of buffer solution (in general, the resistance depends on ion concentrations), most of the applied gate voltage drops on a double layer formed in 2D/electrolyte interface known as an electrical double layer (EDL)[69-71]. The EDL induces a capacitance (called $C_{dl}$) that absorbs the opposite charges inside the channel in which the electrolyte acts as a dielectric [72]. Since the EDL thickness is about a few nanometers, the induced capacitance will be several orders of magnitude higher than the induced back-gate capacitance [59]. Therefore, the conductivity of the bio-FET in a wet state is more sensitive than in a dry state and can be modulated with small charge perturbations[59]. In addition to $C_{dl}$, the applied gate voltage is dropped across another capacitor known as quantum capacitance ($C_Q$)[73]. The $C_{dl}$ value is calculated according to the equation, $C_{dl}=\varepsilon_r\varepsilon_0 A/\lambda_D$, where $\varepsilon_r$ is the relative permittivity of the electrolyte, $\varepsilon_0$ is vacuum permittivity, A is the common area of 2D channel/electrolyte interface, and $\lambda_D$ is the Debye length[73]. $C_Q$ originates from changes in total charge to chemical potential (Fermi level, $E_F$) of 2D material and is defines as $C_Q=dQ/dE_F$[74]. Both parts contribute to the total capacitance as follow[73, 74]:

$$\frac{1}{C_{tot}} = \frac{1}{C_{dl}} + \frac{1}{C_Q} \qquad (1)$$

2D materials have a higher quantum capacitance compared to other materials[74]. Indeed, a small change in their density of state (DOS) results in a significant change in their Fermi level[74]. In general, $C_{dl}$ is usually one order of magnitude larger than $C_Q$, which leads to a dominant contribution of $C_Q$ in the total capacitance[75]. Therefore, any accumulation of biotargets on the surface of the 2D channel leads to a considerable change in its Fermi level and subsequently changes the total capacitance. Hence, it results in the detection of the biotargets by modulation of the conductivity.

In solution-gated bio FETs, Debye length ($\lambda_D$) plays a key role in determining the limit of detection (LOD)[76]. The LOD refers to the lowest detectable amount of biotarget measured by the biosensor and expresses in terms of concentration (or amount). The Debye length is a distance where a net charge is screened to 1/e of its maximum value. The accumulation of buffer ions around the biomolecules is responsible for this screening. The $\lambda_D$ is defined as[77]:

$$\lambda_D = \sqrt{\frac{\varepsilon_r \varepsilon_0 K_B T}{2 N_A e^2 I}} \qquad (2)$$

Where $\varepsilon_0 \varepsilon_r$ is dielectric constant of electrolyte, $K_B T$ is thermal energy, $N_A$ is Avogadro number, e is electron charge, and I is the ionic strength of solution. Accordingly, a solution with a higher ionic concentration has a shorter screening length while longer $\lambda_D$ occurs at lower ionic concentrations[78].

A variety of buffer solutions such as phosphate-buffered saline (PBS), saline-sodium citrate (SSC), sodium fluoride (NaF), water, and Potassium chloride (KCl) are employed as an electrolyte in bio-FETs[79-82]. Among them, PBS is widely used because it is not only biocompatible but also has ionic concentrations similar to the human body. However, solutions with different ionic concentrations provide different Debye lengths [83]. Panel (**f**) to (**h**) of **Figure 2** demonstrate schematic illustrations of the detection process in three PBS solutions with different ionic concentrations. Accordingly, phosphate buffer concentrations of 0.01 M (1×PBS), 0.1 M (0.1×PBS), and 0.01 M (0.01×PBS) have $\lambda_D$ of approximately 0.7, 2.4, and 7.4 nm, respectively[83]. As a result, a buffer solution with a longer $\lambda_D$ detects biotargets more effectively since net charges screen over a longer distance[83]. It is noteworthy that the net charge of a specific biomolecule can be positive or negative depending on its isoelectric point (PI) value relative to the pH of the buffer solution[84, 85]. If PI is larger than the pH of the solution, $H^+$ ions protonate the carbohydrates and neutralize them, leaving only positive amine groups[85]. As a result, the biomolecule possesses a positive net charge in the corresponding solution.

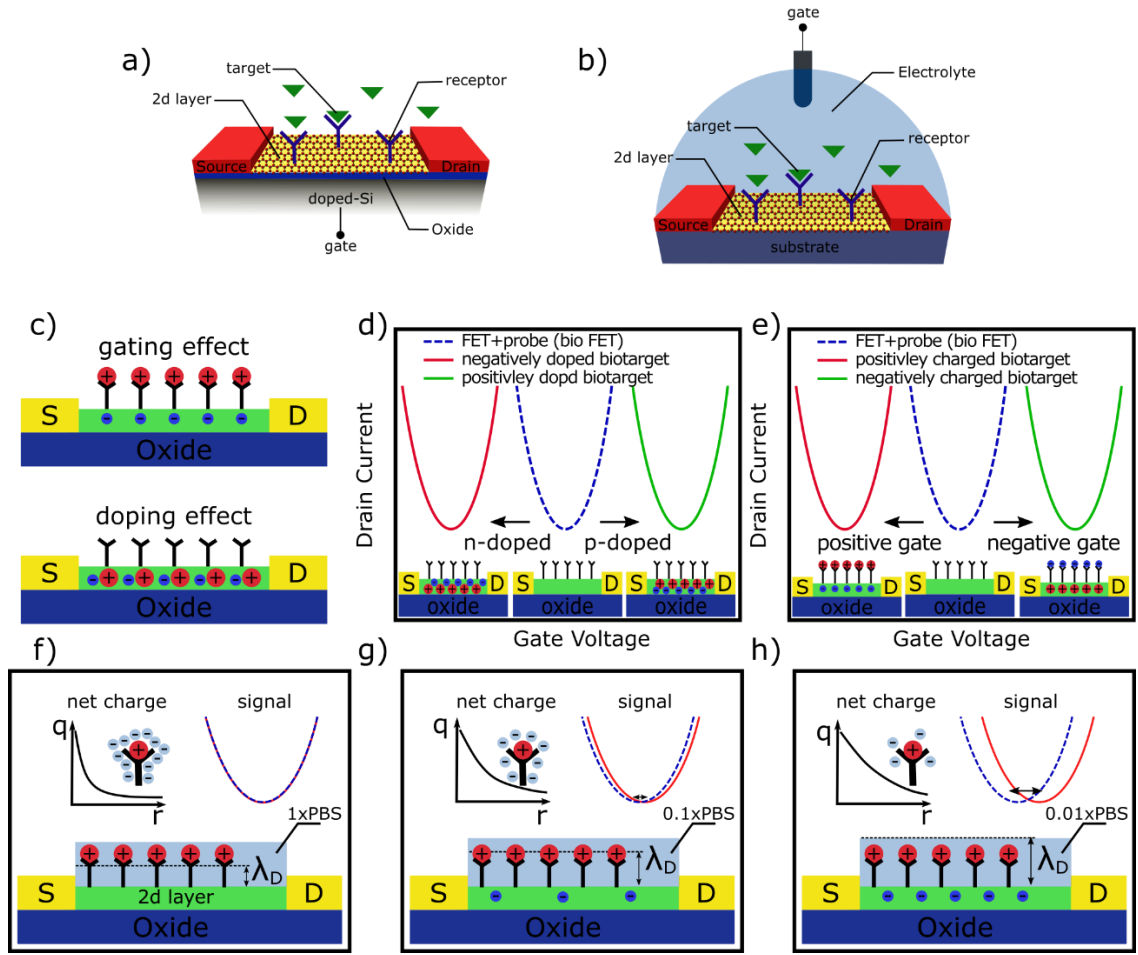

**Figure 2.** Schematic illustration of the 2D based **a)** back-gated and **b)** solution-gated bio-FET configurations. **c)** Doping and gating effects. Change of $I_{ds}$-$V_g$ characteristics upon **d)** doping effect and **e)** gating effect. Solution-gated bio-FETs containing different PBS concentrations with corresponding Debye lengths. **f)** 1×PBS. **g)** 0.1×PBS. **h)** 0.01×PBS.

### 1.1. *2D based FET aptasensors*

Aptamer, a single-stranded DNA or RNA molecule, is artificially generated by an in vitro selection technique called systematic evolution of ligands by exponential enrichment (SELEX) method[86]. In this method, large randomized oligonucleotide sequences are synthesized and a specific aptamer is designed for the proposed biotargets via an iterative process in which aptamers binding to the specific biotargets are enriched. As a result, high-affinity aptamers with high specificity can be made using these complementary sequences[86]. Simple synthesis, low-cost production, thermal stability, and low immunological activity are some advantages of aptamers[86-88]. The discovery of aptamer brings new opportunities in the development of biosensors to detect various biotargets with higher selectivity and sensitivity. FET aptasensors can detect various biomolecules by using specific aptamers, which are produced by the SELEX process and commercially are available.

With the increase in our knowledge of 2D materials, 2D bio-FETs have opened up new doors in introducing better-performing aptasensors. In this case, the specific aptamers are immobilized on the surface of the 2D channel through appropriate linkers. Linker molecules usually have pyrene functional groups that bind to the surface of the channel through π-π, van der Waals, and other interactions[89]. Metal nanoparticles such as AuNPs and PtNPs can also be used as linkers instead of organic molecules[64, 89]. In this case, the charge exchange between aptamer and nanoparticles is responsible for the binding. After immobilization of the aptamer probe, its specific target is introduced into the bio-FET. Depending on biotarget types, different interactions may occur between aptamer and target (such as electrostatic interactions, hydrogen bonding, and so on) where the corresponding binding results in a change of the drain current associating with detection of the biotarget[90].

**Figure 3a** shows a schematic illustration of a FET aptasensor configuration. In addition to organic linkers, metal nanoparticles such as AuNPs can also use as linkers (**Figure 3b**). The surface functionalization and aptamer immobilization can characterize with various analyses including atomic force microscopy (AFM), Raman spectroscopy, electrical characteristics, X-ray photoelectron spectroscopy (XPS), Energy-dispersive X-ray spectroscopy (EDS), and fluorescence microscopy methods[65, 67, 87, 88, 91]. In AFM analysis, thickness is measured after surface modifications[67, 88]. **Figure 3c** shows the measured thicknesses of single-layer graphene with and without functionalization after aptamer immobilizations[91]. It is observed that the thickness of the non-functionalized graphene remains at 0.3 nm while it increases to 2.5 nm in the case of functionalized graphene which refers to the successful immobilization of aptamers. In Raman analysis, coupling of pyrene groups of linkers with 2D material can affect both the position and intensity of oscillation modes of the 2D layer[65, 91, 92]. In the case of electrical measurements, linkers and aptamers change the conductivity or the transfer curve of the 2D FET due to their doping or gating effects[29, 65, 67, 92]. Moreover, they contain elements such as nitrogen and phosphorus, which XPS, EDS and fluorescence analyses can be employed to identify their presence [87, 91].

For applying gate voltages, different configurations have been introduced in 2D based FET aptasensors. For example, gate electrode can be placed planar next to the source and drain electrodes (**Figure 3d**)[93] or immersed in a buffer solution[63, 65, 67, 88, 91, 92]. In dry state bio-FETs, it is usually applied through a back-gated silicon[94].

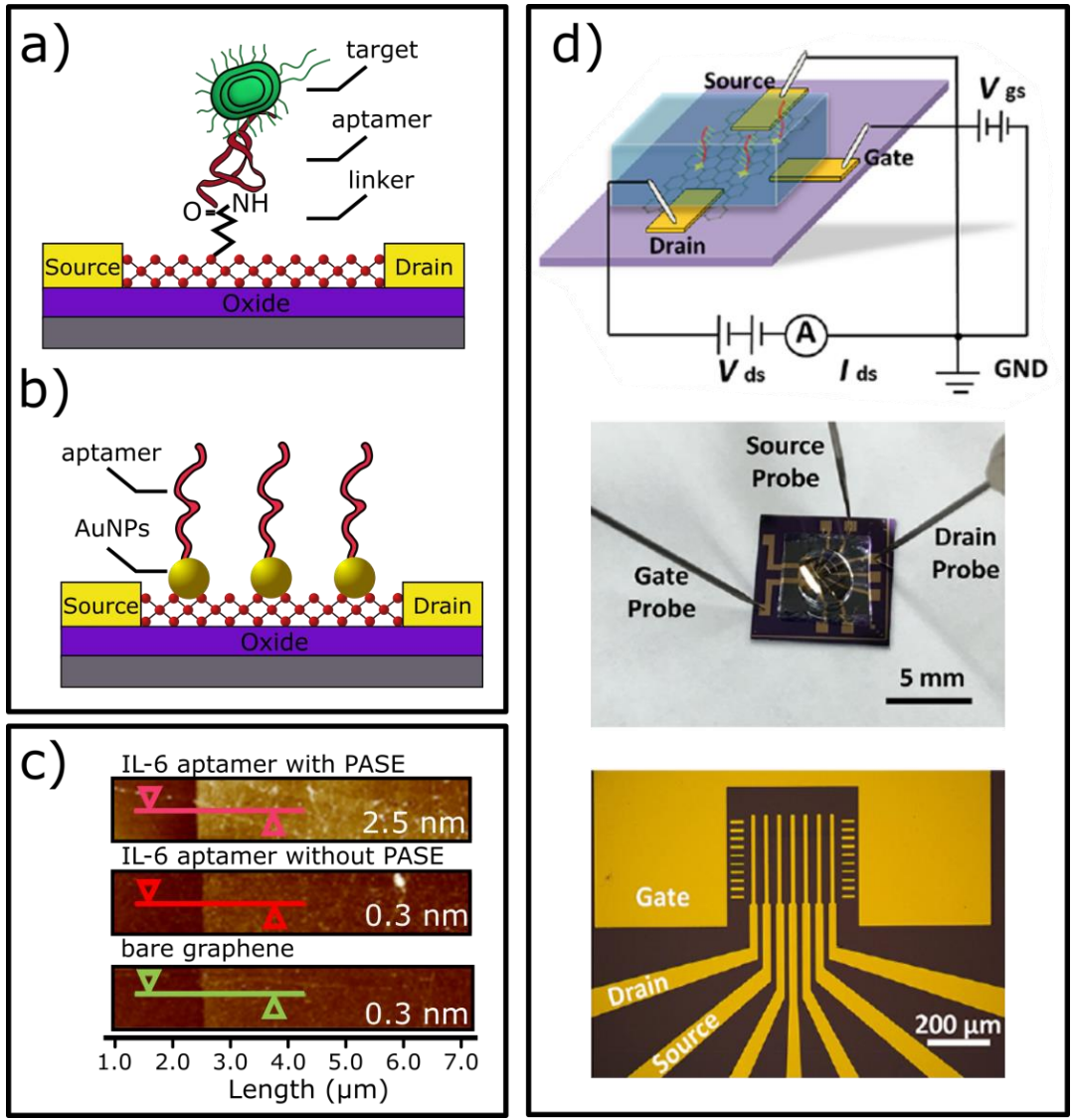

**Figure 3.** 2D-based FET aptasensors: **a)** Schematic configuration of a conventional aptamer bio-FET, and **b)** aptamer immobilizations via AuNPs linker. **c)** AFM measurements of aptamer immobilizations on graphene [91]. **d)** Schematic configuration of an aptamer bio-FET with a planar gate electrode (top), a photograph of the designed bio-FETs (center) and an optical image of the planar-patterned source, drain and gate electrodes (bottom) [93].

Among the large number of reported 2D-based FET aptasensors, special attention has been paid to graphene and Reduced graphene oxide (RGO), thanks to their available π-π interactions, large effective surface area, and environment-dependent electrical transports[29, 67, 87, 88, 93, 95]. Wu et al. reported a graphene-based FET (GFET) capable of detecting Escherichia coli (E.coli) with a limit of detection (LOD) of $10^2$ CFU/mL[67]. In this work, the specific aptamers are directly tagged with pyrene groups and immobilized on graphene. The binding of negatively charged E.coli bacteria to aptamer right-shifts the $\sigma_{min}$, like an electrostatic gate as shown in **Figure 4a**. Ghosh et al introduce a GFET to detect lysozyme in a range of 1 μM to 100 nM[63]. Graphene surface is functionalized with 1-Pyrenebutyric acid N-hydroxysuccinimide ester (PASE) linkers through π-π interactions. Aptamers are then immobilized using amide bond formation with the –NHS of PASE[96]. The electron-rich nucleotide bases of aptamer lead to an n-doping effect while additional of the positively charged lysozyme imposes the p-doping effect and shifts the Dirac point to the positive gate voltage (**Figure 4b**). Hao et al. reporte a GFET for the detection of insulin in a dynamic range of 100 pM - 1 μM with a LOD of 35 μM[93]. Insulin acts as an n-dopant and denotes electrons to graphene. Farid et al. have observed a similar mechanism in GFET for detection of interferon-gamma (INF-γ) in which pyrene-modified aptamers are directly added to graphene [88]. Attachment of INF-γ to the modified graphene leads to an n-doping effect. In another report by Hao et al., a GFET is introduced to identify interleukin 6 (IL-6) in a range of 1 pM - 100 nM[91]. They employed an electric field between the Ag/AgCl gate electrode and the graphene surface to immobilize the PASE linker, which improved the LOD from 1.22 pM to 618 fM. Wang et al. utilized a GFET to detect tumor necrosis factor α (TNF-α) that plays an important role in cellular signaling[65]. In this report, the gate voltage is applied through a planar electrode and artificial tear is used as an electrolyte. TNF-α is detected through the n-doping effect with a LOD of about 2.75 pM. **Figure 4c** shows the response of the GFET to the various concentrations of TNF-α. Based on the electrostatic gating mechanism, Nekrasov et al., and Yang et al. use GFETs to detect ochratoxin A (OTA) and human epidermal growth factor receptor (HER2) biomolecules, respectively[92, 97]. In the latter, the positively charged HER2 screens the induced negative charge of the aptamer and acts as a positive electrostatic gate that decreases the drain current of the p-type GFET.

besides GFET, RGO bio-FETs (RFET) are widely used to detect various biomolecules. RGO is electrically charged due to its carboxyl groups and defects and can provide improved interaction with targets [87, 95]; [98]. An RFET with AuNPs linkers is used by We et al. to detect Hepatocellular Carcinoma-Derived Microvesicles (HEPG2-MVs) in a dynamic range of 500 pM-50 nM with a LOD of 4.8 pM[87]. MVs are bilayer lipids, released from cell membranes, and can transport DNAs or proteins between cells[99]. HEP2-MVs play an intercellular communication role in HEP2 cancer cells and its identification is critical for cancer diagnosis[100]. The capture of HEP2-MVs by aptamer modified RFET results in the accumulation of negative charges on the RGO surface. In another work that is presented in **Figure 4d**, Kim et al. utilized RFETs to identify negatively charged protective antigens (PA) with a LOD of 12 aM [95].

Molybdenum disulfide ($MoS_2$) based FETs (MFET) are also employed to detect various biotargets based on aptamer probes. Chen et al. detects kanamycin (KAN) antibiotics in a dynamic range of 1 nM to 100 μM with a LOD of about 1.06 nM[94]. Aptamers are immobilized on $MoS_2$ surface via Au-S bonds and in order to increase sensitivity, short complementary strand DNAs (cDNA) are also attached to aptamers. The aptamer/CS complex folds together to form a shorter length of about 5.66 nm, which is less than the Debye length of the electrolyte ($\lambda_D$ ~ 5.88 nm). By introduction of KAN molecules, CS is separated form aptamer and aptamer-KAN binding is formed via a non-covalent binding (**Figure 4e**). This process decreases the drain current of MFET since CS elimination results in less hole density in channel as can be seen in **Figure 4f.** In another work, Park et al. demonstrate that addition of different concentrations of cortisol is associated with a decrease in the drain current of the MFET[101].

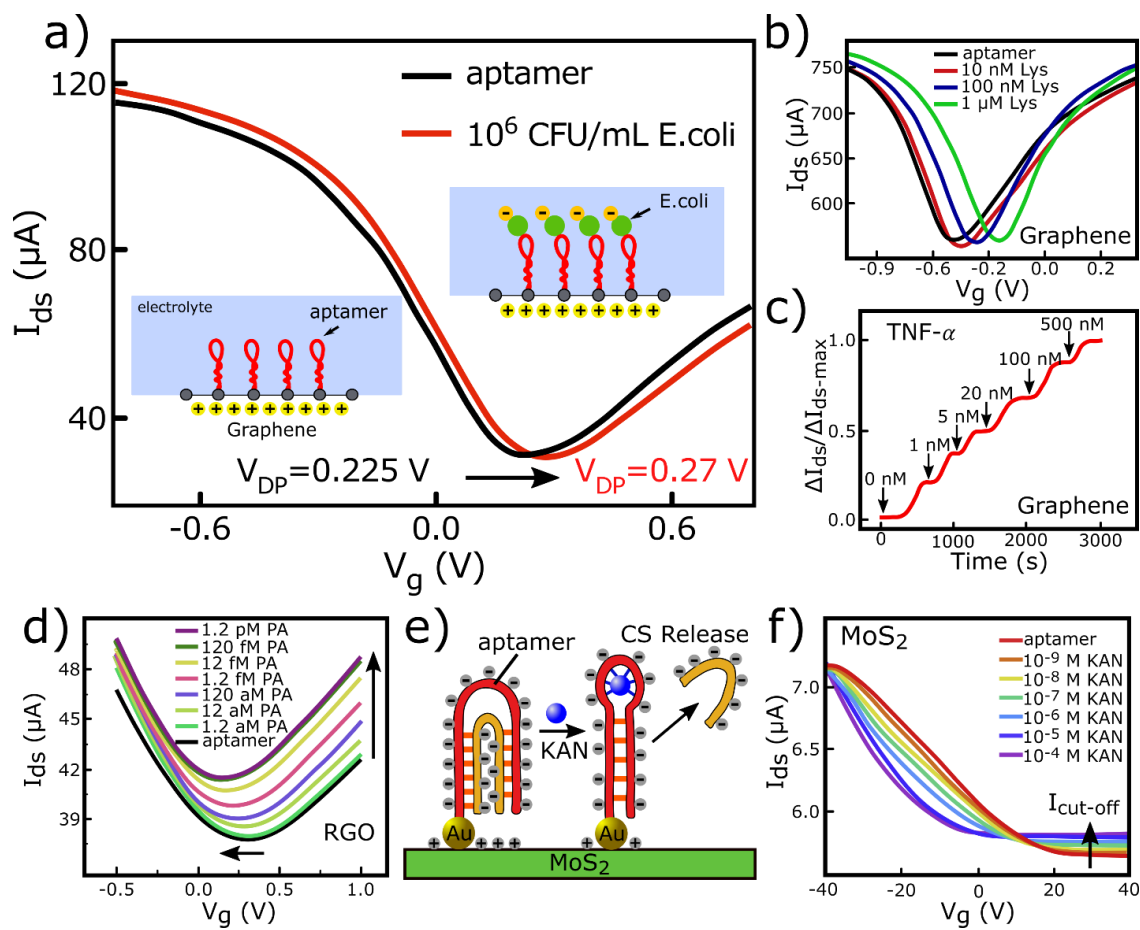

**Figure 4. a)** $I_{ds}$-$V_g$ characteristics of GFET after immobilization of probe aptamer and in response to the E.coli targets[67], **b)** Transfer curve of the GFET under introduction of various concentration of Lys[63]. **c)** Time-resolved measurements of the GFET with different concentration of TNF-γ[65]. **d)** $I_{ds}$-$V_g$ dependency of RFET to different concentration of PAs[95]. **e)** schematic of the sensing mechanism in the MFET in response to KANs, and **j)** corresponding $I_{ds}$-$V_g$ characteristics [94].

Table 1 shows a summary of different RGO, graphene, and MoS$_2$ based FET aptasensors. The table lists the biotarget type, dynamic range, linker, LOD, electrolyte, gate electrode material, and detection mechanism. According to the table, FET aptasensors are capable of detecting various biotargets in a wide dynamic range from aM to μM with a high LOD.

**Table 1.** Performance summary of 2D based FET aptasensors in detection of various biotargets.

| material | target | dynamic range | linker | gate electrode | electrolyte | LOD | sensing mechanism | ref |
|---|---|---|---|---|---|---|---|---|
| RGO | HepG2-Mvs | 500 pM-50 nM | AuNPs | - | 0.01×PBS | 4.8 pM | n-doping | [87] |
| | Protective antigen | 12 aM-120 fM | PASE | Pt wire | 0.001×PBS | 12 aM | n-doping | [95] |
| | HPV-16 E7 | 30-1000 nM | PCA | Ag/AgCl | 10×PBS | 1.75 nM | - | [102] |
| | E.Coli | $10^2$-$10^6$ CFU/mL | Pyrene-tagged aptamer | Ag/AgCl | 0.01×PBS | $10^2$ CFU/ml | gating | [67] |
| | Insulin | 100 pM-1 μM | PASE | Planar Cr/Au | PBS | 35 pM | n-doping | [93] |
| | IFN-γ | 2 nM-100 μM | Pyrene-modified aptamer | Ag/AgCl | PBS | 82.7 pM | n-doping | [88] |
| | IL-6 insulin | 1 pM-50 nM | PASE | Ag/AgCl | 1×PBS | 618 fM 680 fM | doping | [91] |
| | TNF-α | 0.03-500 nM | PASE | Planar Cr/Au | Artificial tear | 2.75 pM | n-doping | [65] |
| Graphene | OTA | 10 pM-1 μM | PASE | Ag/AgCl | PBS | 10 pM | gating | [92] |
| | Lysozyme | 10 nM-1 μM | PASE | Needle | 0.01×PBS | - | p-doping | [63] |
| | HER2 | 54 fM-1 nM | PASE | Ni/Au | 1×PB | 54 fM | gating | [97] |
| | Thr | 1 pM-1 μM | PASE | Planar Au | 0.01×PBS | 2.6 pM | p-doping | [103] |
| | IL-6 | 100 pM- 100 nM | Pyrene-tagged aptamer | Planar Au | 0.01×PBS | 8 pM | n-doping | [104] |
| | CA1 | 0.3 pM- 3 nM | PASE | Pt wire | 0.01×PBS | 70 pM | gating | [105] |
| | IL-6 | 0-100 nM | PASE | back-gated Si | 1×PBS | 10.5 pM | - | [106] |
| MoS$_2$ | KAN | 1 nM- 100 μM | AuNPs | back-gated Si | Dry state | 1.06 nM | p-doping | [94] |
| | Cortisol | 2.7 aM- 0.27 pM | APTES+GA | back-gated Si | Dry state | 2.7 aM | - | [101] |
| | cTn1 | 10 fM-1 nM | APTES | - | 1×PBS | 10 fM | - | [107] |

*1.2. 2D based FET DNA biosensors*

Cost-effective, rapid and highly sensitive DNA detection is of great importance in the diagnosis of genetic diseases, especially in the early detection of cancers[108]. Nucleic-acid diagnostics can play a valuable role in modern medicine by providing comprehensive information for medical experts. However, conventional DNA detection methods, including real-time PCR, molecular cloning, fluorescence assays and other methods are often faced with limitations such as low sensitivity, time-consuming and labor-intensive procedures[109-111]. Label-free FET biosensors, as a promising candidate is highly recommended thanks to their high selectivity, sensitivity, and versatility. Furthermore, their performance can be significantly improved by employing 2D materials as the active channel layers. In FET DNA biosensors, the single-strand DNAs called probe DNAs are immobilized via conjugation reactions between their amine groups with appropriate linkers[108]. In some reports, probe DNA is directly binded to the 2D surface without using linker molecules[62, 112, 113]. Target DNAs are then introduced into the bio-FET in order to be recognized. Due to the degree of genetic similarity, both probe and target DNAs are hybridized through a complementary base pairing and form a single double-stranded molecule[114]. Therefore, the hybridization process results in the generation of an electrical signal in the bio-FET. In addition to DNA, RNA is also used as a probe. DNA backbones are made of deoxyribose, linked together via phosphates, while in PNA, backbones are made of repeating N-(2-aminoethyl)-glycine units linked together by peptide bonds[115].

**Figure 5a** shows a schematic illustration of the sensing procedure in a DNA bio-FET composed of linkers, probe DNA, target DNA and hybridization process. **Figure 5b** displays an AFM image of AuNPs as linkers on a GFET used for immobilizing of the probe DNAs[116]. The addition of linkers and probe DNAs is usually accompanied by an increase in the thickness of the 2D layer, which is one of the methods for characterizing the surface modification. According to **Figure 5c,** adding linkers and probe DNAs increase the thickness of graphene to ~2 and ~4nm, respectively[117]. As a case example, AuNPs linkers are used in an RFET by Cai et al. to immobilize PNA probes in order to detect cDNAs. The hybridization process is associated with n-doping effects in the bio-RFET while attachment of two other types of non-complementary and one-base mismatched DNAs do not show significant impacts on transfer curves of the FETs as shown in **Figure 5d**[118]. In another work, Yin et al. use PtNPs linkers to link probe DNAs on RFET through Pt-S covalent bonds[119]. The LOD of the bio-RFET is reported to be 2.4 nM in a dynamic range of 48 nM to 5 μM.

In addition to RGO, there have been several reports of GFET DNA biosensors. Gao et al. used hairpin as a probe DNA to detect 21-mer in length target DNAs in a dynamic range of 1 fM to 100 pM[66]. A hairpin is a single-strand DNA in form of a double helix structure with an unpaired loop[120]. When target DNA is added, the hairpin is opened and a self-assembly reaction occurs between them in which DNA receives a negative charge and acts as an electrostatic gate (**Figure 5e**). Dong et al. suggest a GFET to detect 12-mer DNA with a LOD of 0.01 nM[62]. The DNA probe is directly immobilized on graphene through non-electrostatic stacking interactions. Zheng et al. immobilized 22-mer PNA probes on a GFET using PASE linkers to detect cDNAs in a dynamic range of 1 fM to 100 pM[121]. Among three different types of complementary, non-complementary, and one-base mismatched DNAs, a considerable drain current change occurs for cDNA thanks to its electron-rich nucleobases. Wang et al. utilized GA cross-linkers to immobilize 42-mer probe DNAs via a covalent bonding of –CH=O to amino-group (-$NH_2$) of DNA[122]. Lin et al. immobilized 12-mer probe DNAs directly on graphene to recognize target DNAs in both dry and wet (1×PBS) states[113]. They also employed Hall-effect measurements to investigate the evolution of carrier densities at different stages of surface modifications. **Figure 5f** shows that the hole density increases in both states for each modification stage, indicating p-doping effects. Xu et al. fabricated a GFET containing a PDMS block to detect 22-mer DNA targets (T20) based on the gating effect[123]. **Figure 5g** shows the photographic image of the introduced biosensors where the corresponding electrical measurements are presented in **Figure 5h**. The length of the probe DNA plays a direct role in determining the LOD of biosensors because it is somehow related to the Debye length. Ping et al. immobilized three different probe DNA lengths (22-mer, 40-mer, and 60-mer) on GFETs using PASE cross-linkers to detect the same target DNAs[117]. Their results show that as the length of cDNA increased, LOD improves from 100 pM to 1 fM.

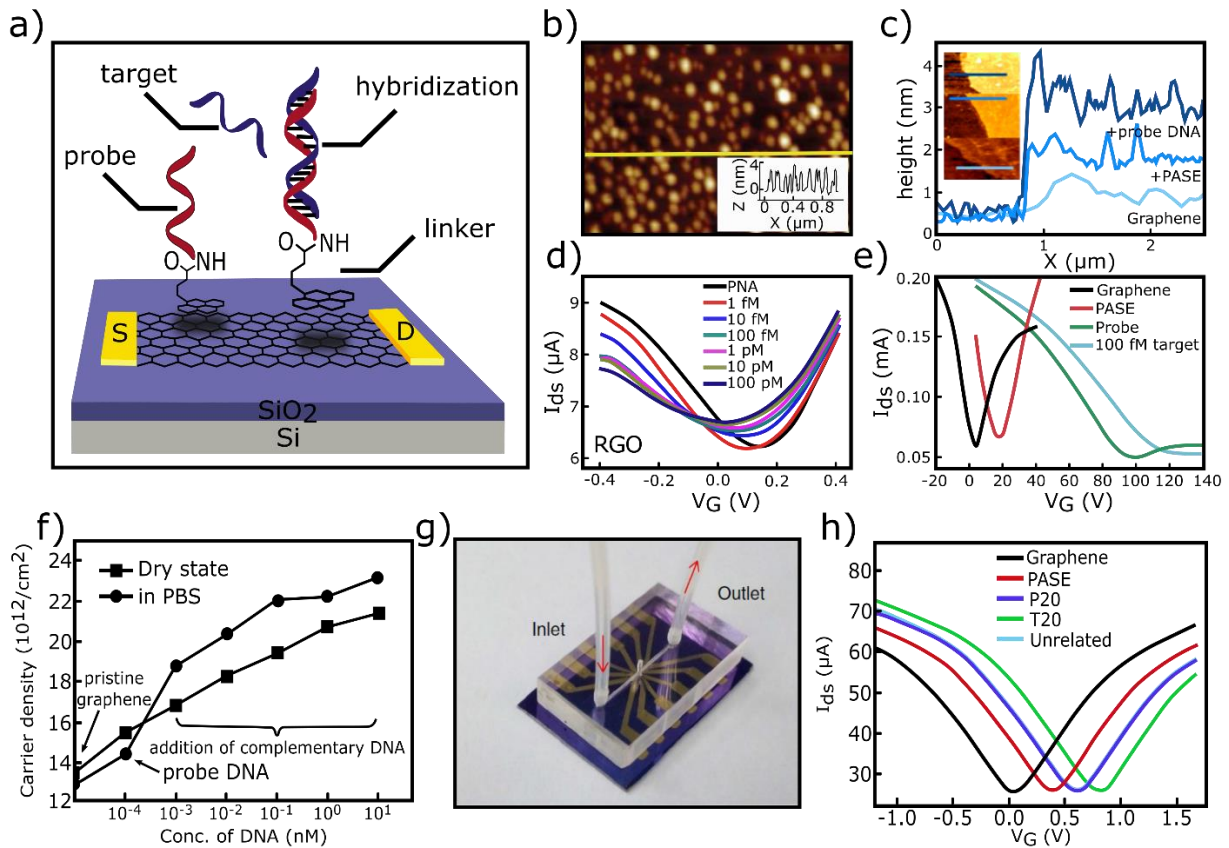

**Figure 5.** GFET DNA biosensors: **a)** Schematic illustration of a conventional DNA bio-FET. **b)** The AFM image of AuNPs on graphene as cross-linkers[116]. **c)** The measured thickness of graphene after modification with PASE linker, and probe DNA[117]. The $I_{ds}$-$V_G$ characteristics of **d)** RFET with adding of various target DNAs[118], and **e)** GFET after modification with PASE linker, probe, and target DNAs in a dry state[66]. **f)** Hole density vs DNA concentrations in a GFET in both dry and wet (1×PBS) states at different stages of modifications[113]. **g)** Image of the DNA based bio-GFET with a PDMS block, and **h)** the corresponding $I_{ds}$-$V_G$ characteristics[123].

Other 2D materials such as MoS$_2$ have been also employed in the 2D based DNA bio-FETs. **Figure 6a** shows a schematic of a bio-MFET introduced by Liu et al[124]. The photograph of the MFET is presented in **Figure 6b**. The surface modifications with AuNPs, probe DNAs, and target DNAs are controlled by Raman analysis as shown in **Figure 6c**. After the attachment of probe DNAs, the A$_{1g}$ peak shifts blue due to its contribution as a p-dopant. The same trend is also observed after DNA hybridizations[124]. Therefore, according to **Figure 6d**, the decrease in drain current (I$_{ds}$) originates from the decrease in the electron density within the MFET channel during DNA hybridizations. Based on the gating mechanism, Lee et al. detect 18-mer probe DNAs as follows: phosphate backbones of the probe DNAs are negatively charged where their accumulation on MoS$_2$ reduces the electrons density within the MFET channel. Hybridizations result in a reduction of the binding interactions of the DNAs with the 2D layer which causes the hybridized DNAs to leave the MoS$_2$ surface. This phenomenon ends with increases of I$_{ds}$ current in the MFET (**Figure 6e**)[112]. Mie et al. immobilize phosphorodiamidate Morpholino oligomer (PMO) as a probe on few-layered MoS$_2$ via PASE cross-linkers[125]. PMO has a similar structure to DNA, except that instead of deoxyribose rings, it has morpholino rings linked together by phosphorodiamidate bonds[126]. In their report, MFET exhibits a p-type behavior and drain current is reduced by the addition of the cDNA. **Figure 6f** shows the current response of the bio-MFET to different targets, with the highest sensitivity to cDNA.

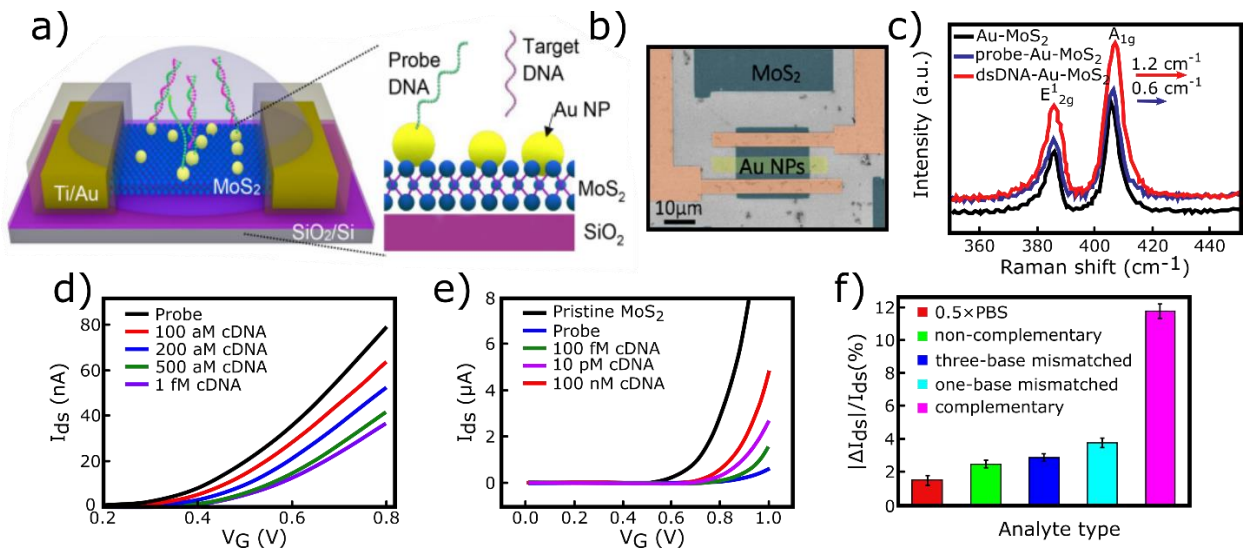

**Figure 6.** Beyond graphene DNA bio-FETs. **a)** Schematic illustration of a MoS$_2$ biosensor, **b)** image of its channel with AuNPs linkers, **c)** Raman analysis of modification states, and **d)** the corresponding I$_{ds}$-V$_G$ curves in response to various concentrations of target DNAs[124]. **e)** Transfer curves of a bio-MFET after immobilizing of probe and target DNAs[112]. **f)** Response of bio-MFET to different target DNAs[125].

A summary of the performance of the DNA bio-FETs based on the various 2D materials and related testing parameters are presented in **Table 2**. These biosensors are able to detect different lengths of target DNAs from 15-mer to more than 60-mer with a wide dynamic range.

**Table 2.** Performance summary of 2D based-FET DNA biosensors in the detection of various target DNAs.

| material | Target DNA | dynamic range | linker | gate electrode | electrolyte | LOD | sensing mechanism | ref |
|---|---|---|---|---|---|---|---|---|
| RGO | 22-mer | 1 fM-100 pM | AuNPs | Ag wire | 1×PBS | 10 fM | n-doping | [118] |
| | 22-mer | 10 fM-1 nM | PASE | Ag wire | 1×PBS | 100 fM | n-doping | [108] |
| | 22-mer | 48 nM-5 μM | PtNPs | Pt wire | PBS | 2.4 nM | n-doping | [119] |
| Graphene | 21-mer | 1 fM-100 pM | PASE | back-gated Si | Dry state | 5 fM | gating | [66] |
| | 12-mer | 0.1-10 nM | Linker free | Ag wire | PBS | 0.01 nM | n-doping | [62] |
| | 22-mer | 1 fM-100 pM | PASE | Ag wire | 0.01×PBS | 10 fM | n-doping | [121] |
| | 42-mer | 1 fM-10 μM | GA | Pt wire | PBS | 10 fM | - | [122] |
| | 12-mer | 1 pM-10 nM | Linker free | Ag wire | 1×PBS | 1 pM | n-doping | [113] |
| | 22-mer | 0.25-10 nM | PASE | Pt wire | 0.01×PBS | 10 pM | gating | [123] |
| | 15-mer | 100 fM-1 nM | PASE | Ag/AgCl | 0.01×PBS | 100 fM | gating | [127] |
| | 22-mer | 100 pM- 100 nM | PASE | Ag/AgCl | 0.1×PBS | 0.1 nM | gating | [128] |
| | 22-mer | 1 fM-100 μM | PASE | back-gated Si | Dry state | 100 pM | gating | [117] |
| | 40-mer | | | | | 100 fM | | |
| | 60-mer | | | | | 1 fM | | |
| | 24-mer | 0.1-10$^4$ nM | AuNPs | back-gated Si | Dry state | 1 nM | gating | [116] |
| | 15-mer | 100 aM-1 pM | PASE | Ag/AgCl | 0.1×PBS | 100 aM | gating | [129] |
| | 22-mer | 600 zM- 60 nM | PASE | Ag/AgCl | 0.1×PBS | 600 zM | gating | [130] |
| Graphene/MoS$_2$ | 15-mer | 10 aM- 100 pM | PASE | Ag/AgCl | 1×PBS | 10 aM | gating | [131] |
| MoS$_2$ | 30-mer | 100 aM-1 fM | AuNPs | Ag wire | 0.1×PBS | 100 aM | p-doping | [124] |
| | 18-mer | 10 fM-100 nM | Linker free | Ag/AgCl | 0.1×PBS | 10 fM | gating | [112] |
| | 22-mer | 10 fM-1 nM | PASE | Ag wire | PBS | 6 fM | - | [125] |

*1.3. 2D based FET immunosensors*

In FET immunosensors, antibodies (Abs) are immobilized as probes on the FET channel by appropriate linkers to detect antigens or biomolecules containing antigens[64, 132-135]. Immunoglobulin, or antibody, is a Y-shaped protein that contains different chains of peptides produced by immune system (plasma) cells[136]. Antibodies can recognize biomolecules called antigens (Ags) or even organic chemicals. Ag is composed of proteins or peptides and can match with binding sites of the specific Ab to form an antibody-antigen complex[137]. When an Ag (target biomolecule) introduces into a bio-FET, its binding to Ab generates a detectable electrical signal and subsequently changes the drain current. This feature is highly efficient in identifying many cancer biomarkers, bacteria, viruses, and proteins[138-144]. In addition, the use of 2D materials as a channel in the bio-FETs dramatically increases their performance efficiency.

**Figure 7a** shows a schematic illustration of a conventional GFET immunosensor with source/drain contacts, graphene as a channel, linkers, probes (Ab), and targets (Ag). Various linkers are employed in the GFET immunosensors such as PASE, metal nanoparticles, 1-pyrenemethylammonium chloride (PyMA), and EDC/NHS[132-134, 142, 143]. **Figure 7b** demonstrates an SEM image of an RFET decorated with PtNPs linkers to detect brain natriuretic peptide (BNP), a useful hormone in diagnosing heart failure, with a LOD of 100 fM[64]. Anti-BNP (probe) is negatively charged at pH of 7.4 since its isoelectric point (PI) is in the range of 6.6 to 7.2. Its immobilization results in an increase of electron density within the channel of RFET. In contrast, BNP Ag (target) is positively charged and its addition acts as a p-dopant as shown in **Figure 7c**. Yu et al. fixed CD63 Abs on a PASE functionalized RGO to detect exosome which contains CD63 Ags on its surface[135]. Exosome has a negative charge and its introduction results in an n-doping of the bio-RFET. **Figure 7d** compares the change of Dirac voltage in response to four different types of biotargets: carcinoembryonic antigen (CEA), alpha-fetoprotein (AFP), bovine serum albumin (BSA) and exosome, where the highest selectivity is for the latest. In another report, Kim et al. used RGO to detect prostate specific antigen-alpha-1-antichymot (PSA-ACT) in two different buffer solutions with pH of 7.4 and 6.2 where PSA-ACT complex is negatively and positively charged at these pH, respectively[132]. According to **Figure 7e**, with the addition of PSA-ACT, Dirac voltage shifts to negative (at pH 7.4) and positive (at pH 6.2) gate voltages. Thakur et al. sensed E.coli cells based on the gating effect in a functionalized RGO[134]. A nanometer-thick $Al_2O_3$ layer is placed between AuNPs linkers and the RGO channel. Adding the negatively charged E.coli cells to positively charged Abs neutralizes Ab net charges and reduces the induced positive gate potential over the channel.

Along with RGO, several reports have been published based on the GFET immunosensors. Seo et al. employed graphene to detect respiratory syndrome coronavirus 2 (SAR-CoV-2) with a detection limit of 0.01 fM to 0.1 pM in PBS buffer and clinical solutions, respectively[144]. **Figure 7f** shows a schematic of the introduced immunosensor. The real-time response of the bio-GFET to different concentrations of SAR-CoV-2 is displayed in **Figure 7g**. Mao et al. functionalized vertically-oriented graphene with AuNPs to detect the most common Ab in human blood called immunoglobulin G (IgG)[133]. Based on it, the addition of IgG is associated with a decrease in $I_{ds}$ current of the bio-GFET. Jung et al. utilized CVD-grown graphene to identify carbohydrate Ags 19-9 (CA 19-9), a marker of pancreatic cancer, at two different pH of 5.8 and 9.4 in 10 μM PBS buffer solutions[140]. CA 19-9 behaves as p-dopant and n-dopant at pH values of 5.8 and 9.4, respectively. Haslam et al. used a GFET to detect human chorionic gonadotropin (hCG) protein in dry state conditions with a LOD of 2 fM [139]. Increased concentration of hCG hormone in the body can be a sign of testicular cancer, prostate cancer, pancreatic cancer or even pregnancy[145]. Kwong et al. identified exosomes in a dynamic range of 1 nM - 0.1 μM with a LOD of 1 nM on the surface of a functionalized graphene[141]. Exosomes contain a great deal of genetic information about cells by which they are secreted, and its identification is helpful in the early detection of cancers[146]. In their work, the anti-CD63 Abs immobilized on graphene with PASE linkers and different concentrations of the exosomes (in PBS solution) were directed to the surface of the bio-FET channel through a PDMS channel. Attachment of exosomes is associated with the electrostatic gating effect. Rajesh et al. immobilized anti-HER3 ScFv on PtNPs decorated graphene to detect HER3 Ags with a detection limit of 1.6 fM[142]. HERs involve information about some tumors such as breast cancer and gastric cancer[142]. Thiol-containing single-chain variable fragment (ScFv) Abs are used as probes, which consists of two light ($V_L$) and heavy ($V_H$) chains that are connected by a peptide linker. 0.1% Tween 20 is also used to block non-specific sites (**Figure 7h**). The electrostatic gating effect originated from the Ag-Ab binding is responsible for the detection of the Ags (**Figure 7i**). Mechanically exfoliated graphenes were employed by Roberts et al. to detect Japanese encephalitis virus (JEV) and avian influenza virus (AIV)[143]. The detection mechanism is described by the reduction of the current through the gating effect. **Figure 7j** shows the response of a bio-GFET fabricated by Gao et al. to identify Borrelia burgdorferi (B. burgdorferi) bacteria, which causes Lyme disease[138]. Two types of anti-IgG and GroEs scFv Abs are used as probes. The LOD of the biosensor is measured to be about 1.3 nM in presence of the anti-IgG probe, which improves to 0.3 pM with GroEs scFv Probe. This difference stems from the fact that scFv Abs (with a length of ~2.2 nm) capture Ags closer to the surface of graphene compared with the anti-IgG (with a length of ~6.9 nm).

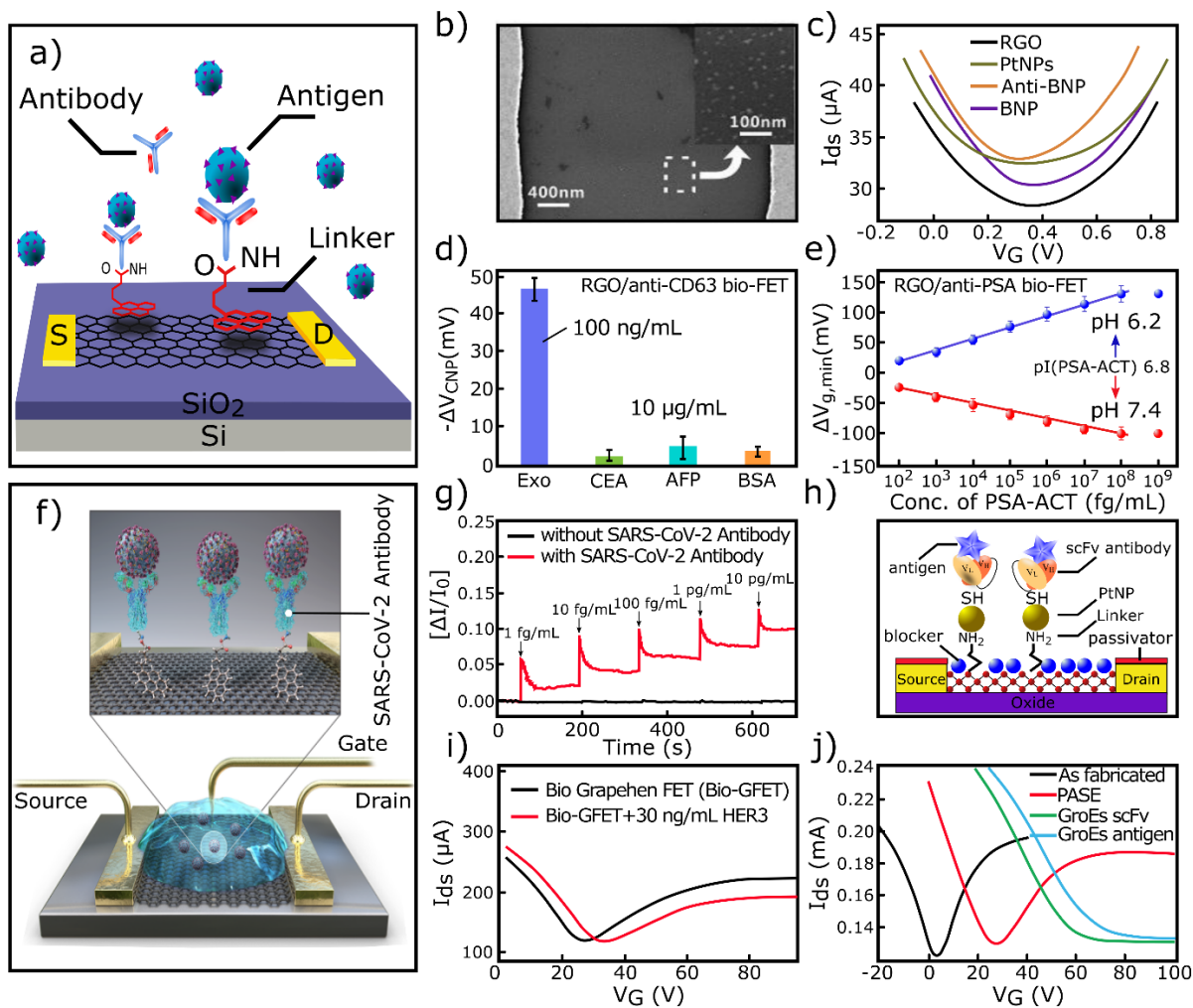

**Figure 7.** GFET immunosensors. **a)** Schematic illustration of a conventional bio-GFET. **b)** SEM image of an RFET decorated with PtNPs linkers and **c)** its $I_{ds}$-$V_g$ characteristics for each stage of modifications[64]. **d)** Change of the Dirac voltage in response to four different biotargets in RGO[135]. **e)** Variation of $\Delta V_{g,min}$ with PSA concentrations in RGO for two testing pH buffer solutions[132]. **f)** Schematic illustration of a bio-GFET for detection of SARS-CoV-2 virus and **g)** real-time sensing in response to SARS-CoV-2 Ags[144]. **h)** Schematic illustration of a bio-GFET with all its components and **i)** corresponding $I_{ds}$-$V_g$ curves in response to 30 ng/mL (0.16 nM) HER3 Ags[142]. **j)** Transfer curve of the graphene after modification with linkers, GroEs scFv antibody and attachment of GroEs antigen[138].

In addition to GFETs, FET immunosensors based on other 2D materials such as TMDs ($MoS_2$, $WSe_2$), back phosphorous (BP), and indium oxide ($In_2O_3$) have been introduced up to date. These 2D materials are generally functionalized by APTES and GA, or in some cases with metal nanoparticles (NPs) and EDC/NHS method, similar to GFETs[147-150]. A schematic illustration of TMD-based FET immunosensors can be seen in **Figure 8a**. Nam et al. embedded $MoS_2$ and $WSe_2$ layers separately on FET channels to detect TNF-α in a 60 fM-6 pM dynamic range (**Figure 8b,c**)[149]. Adding TNF-α to the functionalized FETs reduces the electron densities in both layers referring to p-doping effects (**Figure 8d**). Similarly, Chen et al. sensed TNF-α biomolecules with a LOD of 60 fM based on a MoS$_2$ bio-FET[151]. Ryu et al. functionalized MoS$_2$ FET with APTES and Ga linkers to immobilize anti-IL-1β Abs [150]. In a four-step process called IFDM cycle (Incubation-Flushing-Drying-Measurement), IL-1β is detected in a dynamic range of 1 to 500 fM with a LOD of 1 fM through an electrostatic gate effect. In another report by Park et al., Prostate-Specific Antigen (PSA) biotargets were identified by mechanically exfoliated MoS$_2$ FET with a LOD of 3 fM[152]. **Figure 8e** shows the detection process including surface functionalization with APTES and GA, Anti-PSA immobilization, non-specific sites blocking with casein, and attachment of PSA antigens. Both anti-PSA Abs and PSA Ags are positively charged, and their accumulations on MoS$_2$ induce electrostatic attractions (**Figure 8f**). In another work, Wang et al. applied a microfluidic channel PDMS block to a MoS$_2$ FET to detect PSA Ags[153]. The photograph of the fabricated biosensor can be seen in **Figure 8h**. The $I_{ds}$-t characteristics of the biosensor in response to three different concentrations of PSA are presented in **Figure 8i**. Dalila et al. introduced an MFET immunosensor that detected C-reactive protein (C-RP) Ags with a LOD of 0.06 fM[147]. **Figure 8g** shows the transfer curve of the biosensor in response to different concentrations of the biotargets. The positively charged C-RP Ags increase the density of electrons inside the channel leading to an increase of the drain current. Few layers of black-phosphorous (BP)-bio FET is introduced by Chen et al. to identify IgG[154]. In order to avoid surface passivation, the layer is covered by Al$_2$O$_3$ film. IgG proteins have a negative surface charge and their binding with Abs induce a negative gate voltage on the p-type BP-FET, which leads to an increase in current. **Figure 8j** shows the sensitivity versus target concentrations of the specific probe+IgG Ag, without probe+IgG Ag, and probe+avidin Ag the BP-FET. Accordingly, the highest selectivity is measured for IgG Ags.

Liu et al. used an In$_2$O$_3$ nanoribbon to identify three different biotargets of cardiac troponin I (cTnI), creatine kinase (CK-MB), and B-type natriuretic peptide (BNP), which are commonly associated with heart attack and heart failure[148]. Detection mechanism is reported based on a pH-dependent conductivity called the enzyme-linked immunosorbent assay (ELISA)[155]. This method relies on protonation/deprotonation of OH groups near the FET surface that results in local changes of the gate voltage across the channel. In this report, specific Abs are fixed through EDC/NHS linkers and bind with target Ags. The secondary Abs, which are also specific to the target Ags, bind to the Ags followed by coupling with streptavidins. The other end of the secondary Ab contains biotin where is linked to streptavidin. The biotin-streptavidin complex is one of the preferred arrangements for conjugation with biomolecules[156]. The schematic of the ELISA method is illustrated in **Figure 8k**. Finally, addition of urease increases the pH of solution according to the following reaction:

$$\text{Urea} + 2H_2O + H^+ \rightarrow 2NH^+_4 + HCO^-_3 \tag{3}$$

Accordingly, the protonation of the hydroxyl groups on the surface increases the negatively surface charges and decreases current. The amplification of the detection signal and detection beyond the Debye length limit are advantages of the ELISA method. **Figure 8i** shows the real-time response of the biosensor for three different cTnI concentrations which adding 10 mM urea to the buffer solution results in a reduction of the drain current.

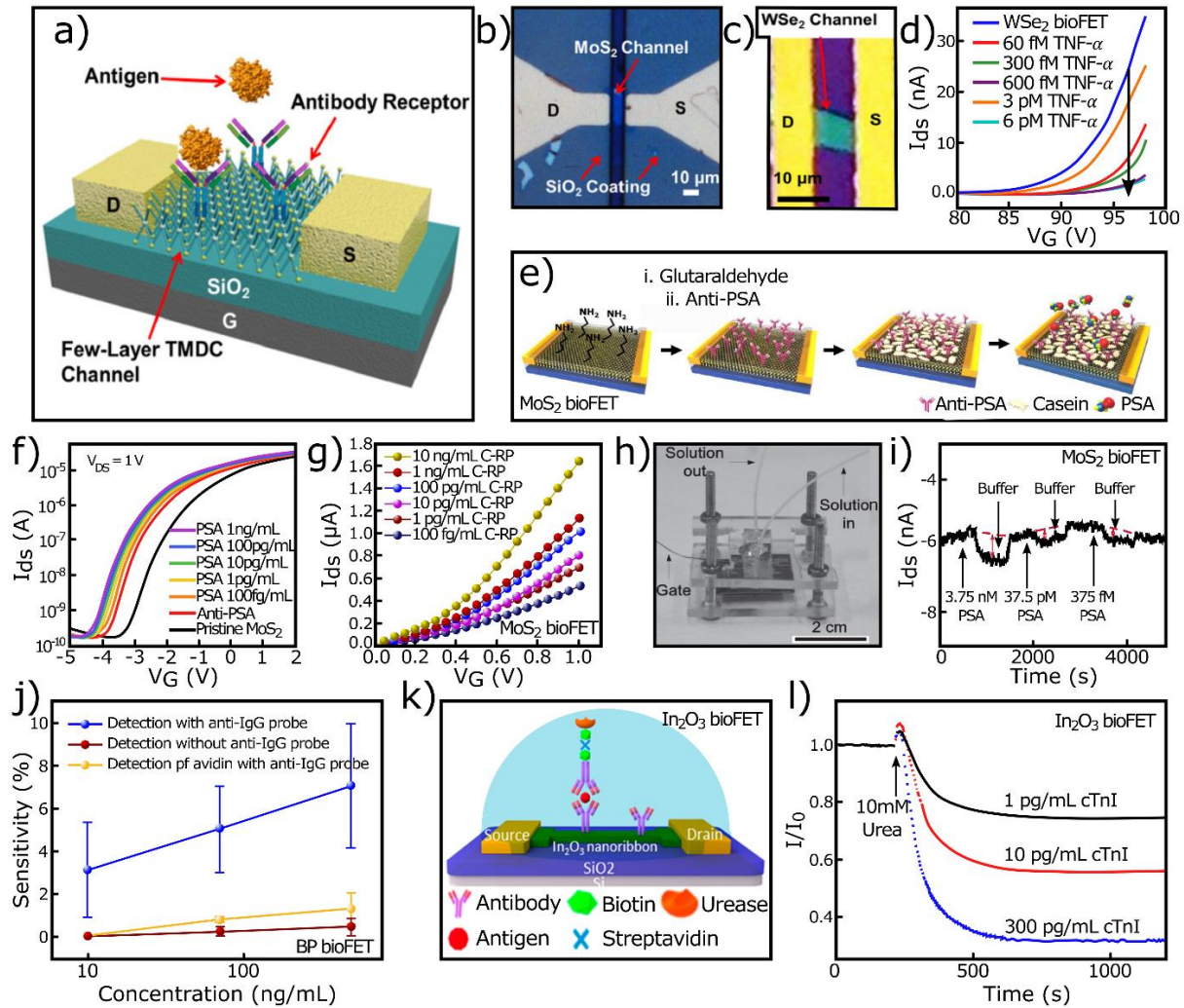

**Figure 8.** Beyond GFET immunosensors. **a)** Schematic illustration of a TMD based bio-FET[149], **b)** image of $MoS_2$ channel, **c)** image of $WSe_2$ channel, and **d)** $I_{ds}$-$V_g$ characteristics of $WSe_2$ bio-FET in response to TNF-α Ags[149]. **e)** Schematic illustration of fabrication stages of a $MoS_2$ bio-FET, and **f)** its $I_{ds}$-$V_g$ curves with addition of PSA Ags[152]. **g)** Transfer curve of a $MoS_2$ bio-FET in response to various concentration of C-RP biotargets[147]. **h)** Image of the

MoS$_2$ bio-FET with PDMS microfluidic block, and **i)** real time sensing to different concentrations of PSA Ags[153]. **j)** Sensitivity versus target concentration in a black phosphorous bio-FET[154]. **k)** Schematic illustration of ELISA detection method in an In$_2$O$_3$ bio-FET, and **l)** real time detection of cTnI Ags with different concentrations [148].

**Table 3** shows a summary of 2D-based FET immunosensors, with corresponding parameters including types of biotargets, linkers, testing conditions, LOD, and the detection mechanism. FET Immunosensors can detect a variety of biomolecules with considerable sensitivity and LOD.

**Table 3.** Performance summary of 2D based FET immunosensors in detection of various Ag targets.

| material | Target antigen | dynamic range | linker | gate electrode | electrolyte | pH | LOD | sensing mechanism | ref |
|---|---|---|---|---|---|---|---|---|---|
| Doped GO | BNP | 10 aM- 1 μM | Linker-free | back-gated Si | Buffer | 7.0 | 10 aM | gating | [157] |
| | exosome | 1 fM- 0.1 pM | PASE | back-gated Si | PBS | 7.4 | 1 fM | n-doping | [135] |
| RGO | PSA-ACT | 1.1 fM- 0.01 μM | PASE | pt wire | 0.0001×PBS | 7.4 / 6.2 | 1.1 fM | n-doping p-doping | [132] |
| | E. coli | 10$^3$-10$^5$ CFU/mL | AuNPs+EDC/NHS | back-gated Si | Dry state | - | 10$^3$ CFU/mL | gating | [134] |
| | BNP | 100 fM-1 nM | PtNPs+EDC/NHS | Ag wire | 0.001×PBS | 7.4 | 100 fM | p-doping | [64] |
| Graphene | IgG | 13 pM-0.1 nM | AuNPs | back-gated Si | Dry state | - | 13 pM | - | [133] |
| | CA 19-9 | 0.01-1000 unit/mL | PASE | Ag/AgCl | 0.001×PBS | 9.4 / 5.8 | 0.01 unit/mL | n-doping p-doping | [140] |
| | HER3 | 1.6 fM- 1.6 mM | PyMA+PtNPs | back-gated Si | 0.1×PBS | 7.3 | 1.6 fM | gating | [142] |
| | hCG | 2 fM- 20 pM | PASE | back-gated Si | Dry state | - | 2 fM | - | [139] |
| | exosome | 1 nM- 0.1 μM | PASE | back-gated Si | PBS | 7.0 | 1 nM | gating | [141] |
| | JEV AIV | 1 fM-1 μM | EDC/NHS | back-gated Si | 5×PBS (50 mM) | 7.4 | 1 fM 10 fM | gating | [143] |
| | B. burgdorferi | 0.3 pM- 3 nM | PASE | back-gated Si | Dry state | - | 0.3 pM | gating | [138] |
| | SARS-CoV-2 | 0.01 fM- 0.1 pM | PASE | - | PBS | 7.4 | 0.01 fM | - | [144] |
| | Clusterin | 10 fM- 1.6 pM | Pyr-NHS | back-gated Si | Dry state | - | 5 fM | - | [158] |
| MoS$_2$ WSe$_2$ | TNF-α | 60 fM-6 pM | APTES+GA | back-gated Si | Dry state | - | 60 fM | p-doping | [149] |
| WSe$_2$ | SARS-CoV-2 | 0.3 fM-0.1 nM | EDC/NHS | back-gated Si | 0.01×PBS | - | 0.3 fM | gating | [159] |
| | PSA | 0.3 fM- 0.03 nM | APTES | back-gated Si | Dry state | - | 0.3 fM | gating | [160] |
| | PSA | 375 fM-3.75 nM | APTES+GA | Ag/AgCl | 0.01×PBS | 6.68 | 375 fM | gating | [153] |
| | PSA | 3 fM- 30 pM | APTES+GA | back-gated Si | Dry state | - | 3 fM | gating | [152] |
| MoS$_2$ | C-RP | 0.8 fM- 8 pM | AuNPs | back-gated Si | PBS | 7.4 | 0.06 fM | gating | [147] |
| | TNF-α | 60 pM-60 fM | APTES+GA | back-gated Si | Dry state | - | 60 fM | - | [151] |
| | ConA | 1 mM- 100 nM | AuNPs | back-gated Si | Dry state | - | 105 nM | gating | [161] |
| | IL-1β | 1-500 fM | APTES+GA | back-gated Si | Dry state | - | 1 fM | gating | [150] |
| BP | IgG | 60 pM- 3 nM | AuNPs+GA | back-gated Si | 0.01×PBS | - | 60 pM | gating | [154] |
| In$_2$O$_3$ | cTnI CK-MB BNP | 40 fM- 12 pM 1 fM- 0.03 nM 3.3- 30 pM | EDC/NHS | Ag/AgCl | 0.01×PBS | - | 40 fM 1 fM 3.3 pM | gating | [148] |

*1.4. 2D based FET enzymatic biosensor*

The enzymatic reaction is responsible for generating a detectable electrical signal between probe enzyme and target biomolecules[162]. Enzymes are proteins that usually act as catalytic molecules where their product of reaction affects the conductivity of the bio-FETs[81, 163, 164]. Similar to other FET biosensors, enzymes are immobilized on the surface of a 2D channel as probes through appropriate linkers.

**Figure 9a** shows a schematic of a GFET enzymatic biosensor[165]. **Figure 9b** demonstrates a GFET enzymatic biosensor introduced by Thanh et al to detect 1-naphthyl-N-methyl carbamate (carbaryl)[82]. Carbaryl is a chemical used to control insects in agricultural products, and excessive accumulation of it in food or water can lead to human poisoning[82]. To recognize carbaryl, first the urease enzymes are immobilized on the surface of graphene, and a mixture of urea/water is added to the biosensor to initiate the following reaction:

Urea+$3H_2O$+urease → $CO_2$+$2NH^+_4$+$2OH^-$

In the presence of urea and enzymes, $NH^+_4$ ions are produced by the hydrolysis of urea, leading to n-doping effects. This hydrolysis process is restricted by the introduction of carbaryl and results in less accumulation of $NH^+_4$ ions on graphene and right-shifts the Dirac point. Kwak et al. employed CVD-grown graphene to detect glucose in a range of 3.1-10.9 mM with a detection limit of 3.3 mM[165]. Glucose oxidase (GOx) is immobilized on the surface of graphene as a probe via PASE linkers. Glucose is catalyzed based on the enzymatic reaction by GOx, in which its reaction products are gluconic acid and $H_2O_2$. The increase of glucose concentration leads to more production of $H_2O_2$, which results in the n-doping effect. **Figure 9c** shows the $I_{ds}$-$V_g$ characteristic of the bio-FET in response to different concentrations of glucose[165]. The corresponding schematic of this biosensor is also presented in **Figure 9a**. In another work, CVD-grown graphene was reported by Zhang et al. to detect glucose with a detection limit of 0.5 μM in a dynamic range of 0.5 μM-1 mM[166]. They used PtNPs as linkers to fix the GOx. Similarly, the glucose detection process is associated with the release of $H_2O_2$ and an increase in the drain current.

In addition to graphene, RGO has also been used as FETs in enzymatic biosensors. Park et al. used 4-(4,6-dimethoxy-1,3,5-triazine-2-yl)-4-methyl-morpholinium chloride (DMT-MM) linkers to immobilize GOx on RGO to detect glucose with a detection limit of 1 nM[164]. Piccinini et al. functionalized RGO with sodium 1-pyrenesulfonate (SPS) linkers followed by assemble of urease-polyethyleneimine (PEI) layers as probes to detect urea[81]. This enzyme accelerates the hydrolysis process of urea to produce $NH^+_4$, $HCO^{-2}$, and $OH^-$ ions which locally change the pH of the solution. **Figure 9d** demonstrates the real-time response of the bio-FET to different concentrations of urea[81]. As the urea concentration increases, more charge carriers accumulate on the channel surface, which increases the drain current. Using a PASE-modified RGO, Chae et al. identified acetylcholine (ACh) neurotransmitter with a LOD of 1 μM[163]. In this work, acetylcholinesterase (AChE) was used as a probe and Ach-AChE enzymatic reaction leads to the accumulation of $H^+$ protons on the surface of RGO (**Figure 9e**).

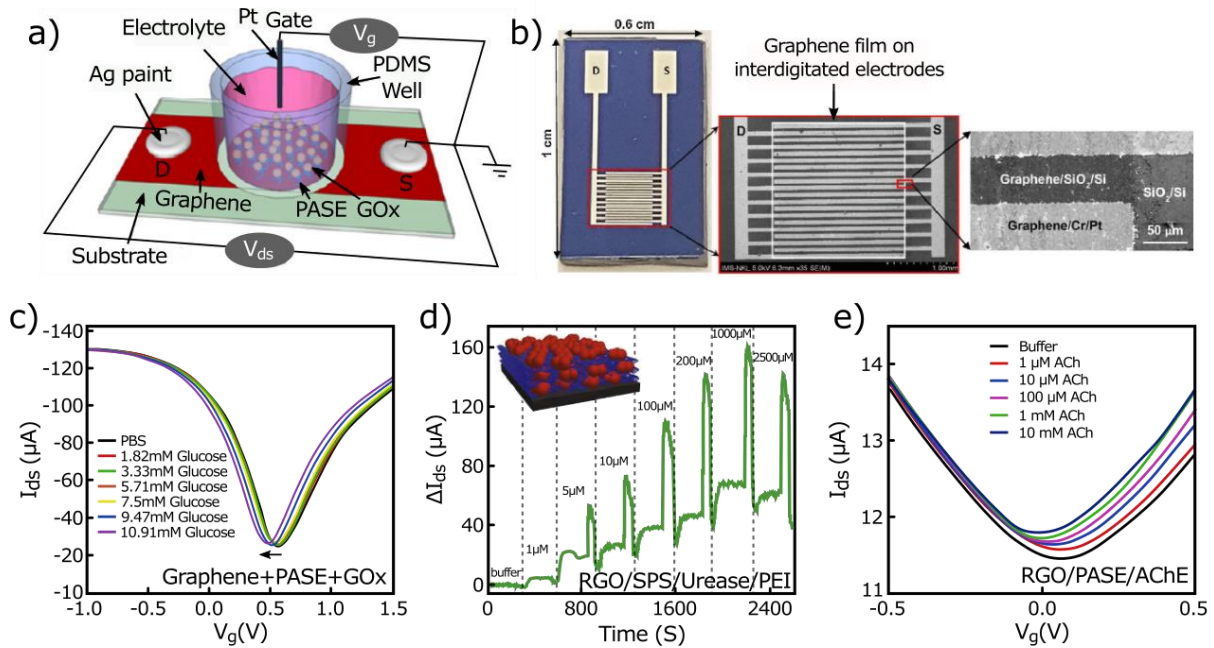

**Figure 9.** GFET enzymatic biosensors. **a)** Schematic illustration of a conventional bio-GFET[165]. **b)** Optical and SEM image of interdigitated electrodes containing graphene as an enzymatic bio-FETs[82]. **c)** $I_{ds}$-$V_g$ characteristics of the GFET biosensor upon different concentrations of glucose[165]. **d)** Real-time response of the RGO biosensor to various inlet urea concentrations from 1 μM to 2500 μM[81]. **e)** Transfer curve of the RGO biosensor in response to different concentrations of Ach biomolecules[163].

Enzymatic biosensors based on other 2D materials such as $MoS_2$ and $WSe_2$ have been also introduced up to date. Lee et al. employed mechanically exfoliated $WSe_2$ few layers to detect glucose in a range of 1 to 10 mM, with the help of APTES+GA and GOx as linkers and probes, respectively[167]. GA cross-linker is used as a middle-linker to facilitate an effective immobilization of GOx on $WSe_2$. **Figure 10a** presents a schematic of the biosensor, the corresponding configuration of the linkers and GOx, and photograph of the FET channel. Upon the enzymatic reactions, $H^+$ ions and electrons are generated (**Figure 10b**) which electrons are directly transferred into the channel causing an n-doping effect as shown in **Figure 10c**. Shan et al. attached GOx directly on $MoS_2$ where detection of glucose was done by a similar procedure[168]. According to **Figure 10d**, as the glucose concentration increases, the drain current increases since more electrons transfer into the $MoS_2$.

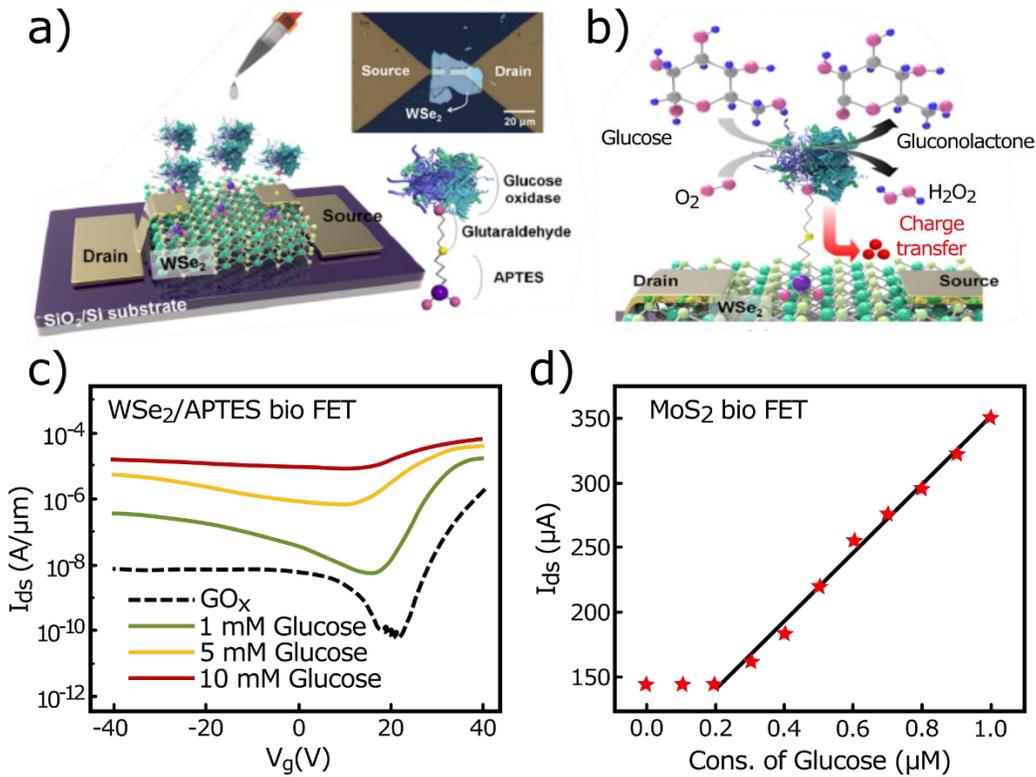

**Figure 10.** Beyond GFET enzymatic biosensors. **a)** Schematic illustration of a WSe$_2$ bio-FET with corresponding molecular configurations and image of the FET, **b)** Detection mechanism of the glucose near surface of the WSe$_2$ bio-FET, and **c)** its $I_{ds}$-$V_g$ characteristics in response to three different glucose concentrations (1, 5, 10 mM)[167]. **d)** $I_{ds}$ versus glucose concentrations in a MoS$_2$ bio-FET[168]. **e)** SEM image of the In$_2$O$_3$ bio-FET with a planar gate electrode, and **f)** real-time response to different concentrations of glucose from 10 nM to 1 mM[169].

**Table 4** summarizes performance of the 2D-based FET enzymatic biosensor in identifying different biomolecules with relevant parameters.

**Table 4.** Performance summary of 2D based FET enzymatic biosensors in detection of various bio targets.

| material | Target antigen | dynamic range | linker | gate electrode | electrolyte | LOD | sensing mechanism | ref |
|---|---|---|---|---|---|---|---|---|
| RGO | glucose | 1 nM-100 mM | DMT-MM | Ag/AgCl | PBS | 1 nM | gating | [164] |
|  | urea | 1-1000 μM | SPS | - | 10 mM KCl | 1 μM | gating | [81] |
|  | ACh | 1 μM- 10 mM | PASE | Ag/AgCl | 0.1 mM PBS | 1 μM | gating | [163] |
| Graphene | carbaryl | 1.2 μM- 1.2 mM | GA | Ag/AgCl | Urea/water | 0.1 fM | gating | [82] |
|  | glucose | 3.3-10.9 mM | PASE | Pt wire | PBS | 3.3 mM | n-doping | [165] |
| | glucose | 0.5 μM-1 mM | PtNPs | sensor | PBS | 0.5 μM | - | [166] |
| | Ach | 5-1000 μM | PABA | Ag/AgCl | KCl (10 mM) + HEPES (0.1 mM) | 2.3 μM | n-doping | [170] |
| | Lactose | 1 aM- 1 nM | AuNPs | Ag wire | 0.01×PBS | 200 Am | n-doping | [171] |
| $MoS_2$ | Glucose | 300 nM-30 mM | Linker free | back-gated Si | 0.1 mol/L PBS | 300 nM | n-doping | [168] |
| $WSe_2$ | Glucose | 1-10 mM | APTES | back-gated Si | Dry state | - | n-doping | [167] |
| $In_2O_3$ | Glucose | 10 nM-1 mM | Chitosan | Planar Au | 0.01×PBS | 10 nM | gating | [169] |

## 2. Bio-medical applications of FET biosensors

Bio-clinical diagnosis offers an innovative approach in monitoring human health status in everyday life. Most conventional methods depend on blood sampling because it provides rich information about blood urea, blood ions, glucose, and other parameters. However, this method is painful, time-consuming, and expensive. As a result, modern medicine desperately needs new solutions.

In addition to blood, the human body can produce other biofluids as clinical samples for disease detection. The human body secretes various biofluids such as tear, sweat, urine, saliva, and serum. All of these biofluids are evaluable as clinical samples for biosensing purposes. Salts, proteins, enzymes, and antibodies are found in the tear. In case of saliva, it contains various cells, enzymes, ions, and antibodies. Urine has glucose, proteins, cells, and bacteria. Sweat is composed of salt, proteins, bacteria, urea, ions, and other minerals and acids. Blood serum contains a variety of biomarkers, lipids, and salt. Therefore, FET biosensors have the advantage of being selective to detect bio targets via all human biofluids. Moreover, it can play a significant role in the future of bio-sensing technology by providing cheap, label-free, high-sensitive, high-selective, and real-time testing.

Up to now, most 2D-based bio-FET are tested in aqueous environments such as PBS at a pH value of 7.4. However, the term "point of care" for these types of biosensors can come true when they could operate accurately in real (clinical) samples. Fortunately, with the dramatic advancement of 2D materials, various 2D-based FET biosensors have gained the ability to real-time monitoring of disease in diluted clinical samples. In fact, human samples are usually diluted with buffer solutions due to their high viscosity[105].

To the best of our knowledge, 2D-based FET aptasensors have been employed to identify bio targets in different real samples such as urine, saliva, sweat, and serum. Hao et al. measured insulin in a diluted human urine sample with a LOD of 680 fM[91]. **Figure 11a** shows the change of drain current of the GFET aptasensor upon adding different insulin concentrations in human urine. The Dirac voltage shift of 10 healthy samples (called healthy control) and 10 patient samples (called hepatocellular carcinoma, HCC) that are collected from human serum are compared in **Figure 11b** for an RGO-based aptasensor[87]. Accordingly, the Dirac voltage shift is clearly detectable in patient samples compared to the healthy samples. In addition to urine and serum, other 2D based FET aptasensors have been introduced to diagnose biotargets in saliva[102, 105, 106] and sweat[71, 172].

In the case of 2D-based FET DNA biosensors, they are mostly tested in human serum samples. In fact, the cDNA (or RNA) is spiked into a diluted human serum and then added to the biosensor. **Figure 11c** shows the $I_{ds}$-$V_g$ characteristic of a GFET DNA biosensor in a fresh serum and a serum containing 100 aM and 1 fM target RNAs[129]. Mei et al. demonstrated that the FET channel containing $MoS_2$+probe DNA well responds to a serum with different concentrations of cDNA (**Figure 11d**)[125].

2D-based FET enzymatic biosensors have been successfully tested in different clinical samples such as saliva, serum, sweat, urine, and tear. Liu et al. employed $In_2O_3$ layer to detect glucose with different concentrations in three clinical samples of saliva, artificial tear and artificial sweat[169]. **Figure 11e** shows the sensing response of the introduced FET upon these three samples compared to 0.1×PBS. It is important to note that in addition to blood, glucose can be found in tears, sweat, and saliva but in lower concentrations[169]. Fenoy et al. spiked Ach in a urine sample and evaluated the FET biosensor's response to its different concentrations[170]. As shown in **Figure 11f**, stronger signals appear as the Ach concentration increases from 5 μM to 0.5 mM.

In addition to these types of FET biosensors, various 2D-based FET immunosensors have been reported based on real samples such as a nasopharyngeal swab, serum, and urine to detect disease. Yang et al. used a functionalized $MoS_2$ layer to identify NMP22 and CK8 biotargets as biomarkers of human bladder cancer in urine[173]. **Figure 11g** presents a schematic illustration of the identification process. Seo et al. utilized a GFET immunosensor to detect the SARS-CoV-2 virus in a diluted sample of the nasopharyngeal swabs[144]. **Figure 11h** shows the sensor response to a healthy sample and a patient sample containing the virus. In addition to urine and nasopharyngeal swab, some other 2D-based FET immunosensors have been tested in human serum samples. Chen et al. spiked different concentrations of the Ebola virus in human serum and tracked the change of $I_{ds}$ per Ebola concentration[174].

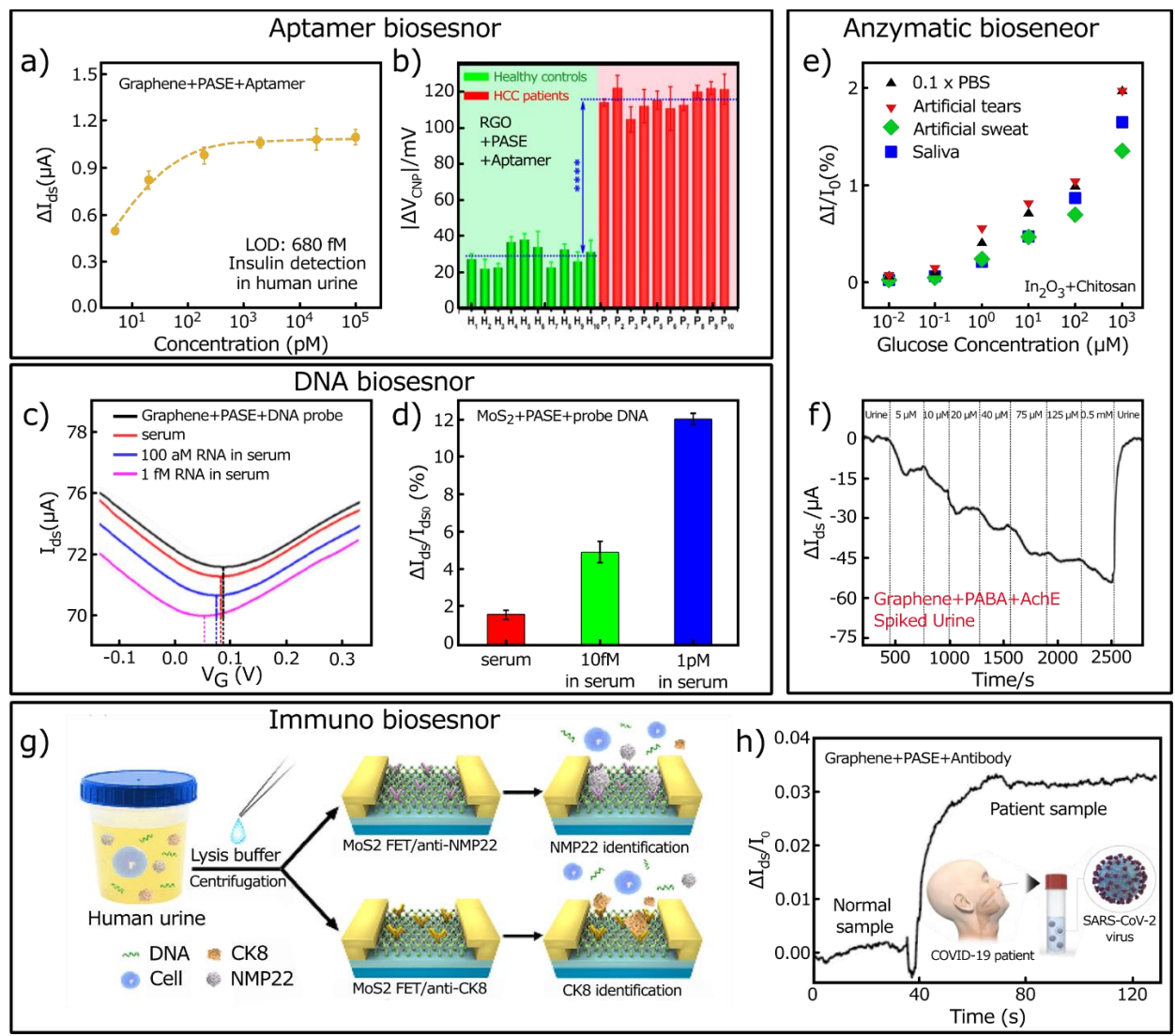

**Figure 11.** Performance of the 2D-based FET biosensors in clinical samples. **a)** GFET aptasensor response to different concentrations of insulin in human urine[91]. **b)** Comparison of RFET aptasensor response to 10 healthy and HCC patient samples collected from human serums[87]. **c)** Dirac voltage shift in a human serum containing 100 aM and 1 fM target RNA in a graphene-based DNA biosensor[129]. **d)** MoS$_2$ based DNA biosensor response to two different concentrations of target DNAs in human serum[125]. **e)** Performance of the In$_2$O$_3$ based enzymatic biosensor in detecting glucose under 0.1×PBS solution, artificial tear, artificial sweat and saliva[169]. **f)** Detection of spiked Ach biotarget in human urine by graphene-based enzymatic biosensor[170]. **g)** Detection process of NMP22 and CK8 biotargets in a MoS$_2$ based FET immunosensor[173]. **h)** Detection of SARS-CoV-2 virus in a patient's diluted nasopharyngeal swab sample via GFET immunosensor[144].

**Table 5** summarizes the performance of the 2D-based FET biosensors to detect biotargets in different human samples of saliva, serum, sweat, urine, and tear. In some of them, the human biofluids directly contain biotargets[102, 105, 107, 173], but in others, certain biotargets are spiked into them[175-178]. These FET biosensors show high accuracy and selectivity in identifying biotargets in real human samples. They can be used to diagnose or monitor diseases in the real world as promising POC devices for modern medicine.

Table 5. Performance of the 2D based FET biosensors in human clinical samples.

| Bio-clinical sample | Bio probe | 2D material | target | LOD | ref |
|---|---|---|---|---|---|
| Saliva and Nasopharyngeal swab* | Aptamer | RGO | HPV-16 E7 | 1.75 nM | [102] |
| | | Graphene | CA1 | - | [105] |
| | | Graphene | IL-6 | 12.2 pM | [106] |
| | Immuno* | Graphene | SARS-CoV-2 | 0.01 fM | [144] |
| | Enzyme | $In_2O_3$ | Glucose | 10 nM | [169] |
| Serum and blood | Aptamer | RGO+CNTs | CA125 | - | [175] |
| | | Graphene | IgE | 47 pM | [178] |
| | | Graphene | ADH | - | [177] |
| | | Graphene | Cortisol | 2.3 pM | [176] |
| | DNA | RGO | RNA | 3.1 pM | [179] |
| | | RGO | miRNA | 10 fM | [118] |
| | | Graphene | RNA | 0.1 aM | [129] |
| | | $MoS_2$ | DNA | 6 fM | [125] |
| | | $MoS_2$ | DNA | 1 fM | [180] |
| | | RGO | Ebola virus | 0.2 fM | [174] |
| | immuno | RGO | CD63 | - | [135] |
| | | Graphene | BNP | 100 fM | [64] |
| | | $MoS_2$ | C-RP | 0.06 fM | [147] |
| | | $MoS_2$ | PSA | 0.03 fM | [181] |
| | Enzyme | RGO | Glucose | 1 nM | [164] |
| Sweat | Aptamer | Graphene | Cortisol | 0.2 nM | [172] |
| | Aptamer | Graphene | INF-γ | 740 fM | [71] |
| | Enzyme | $In_2O_3$ | Glucose | 10 nM | [169] |
| Urine | Aptamer | Graphene | Insulin | 680 fM | [91] |
| | Enzyme | Graphene | Ach | 2.3 μM | [170] |
| | immuno | $MoS_2$ | NMP22 CK8 | 0.027 aM 0.019 aM | [173] |
| Tear | Enzyme | $MoS_2$ | glucose | - | [182] |
| | | $In_2O_3$ | Glucose | 10 nM | [169] |

3. *Conclusion and perspective*

Early diagnosis of diseases is an important factor in reducing disease mortality. Among the various detection methods, FET-based biosensors have gained considerable potential in point of care diagnostics as on-site detection kits due to their high sensitivity, excellent selectivity, and rapid detection. In this review, a comprehensive report on 2D material-based bio-FETs is provided in which their performance in detecting a wide range of biospecies is thoroughly discussed. Commercial availability and high affinity of aptamers result in the implementation of many 2D-based aptamer and DNA FET biosensors, capable of detecting a variety of biomolecules and DNAs. The diversity of 2D-based FET immunosensors is larger than the other three categories and is expected to receive more attention in the future as viral diseases are constantly emerging. Among the four classified types, fewer reports have been published based on 2D-based FET enzymatic biosensors, perhaps due to the complexity of the detection process, in which target is usually measured indirectly based on the enzymatic reaction and ion release.

With recent advances in smart healthcare and wearable biosensors, reliable, fast, and noninvasive point of care devices like bio-FETs are urgently needed to accurately assess human health. As a result, modern medicine will rely more than ever on these point of care biosensors. Despite all these advantages, there are main challenges that need to be addressed to further develop 2D-based bio-FETs: 1) wide varieties of 2D materials have been explored up to now, the most important of which are Mxenes. However, a limited number of 2D materials have been so far introduced as 2D-based bio-FETs and there is a great place for the development of new 2D based bio-FETs. Before that, the biocompatibility of these new 2D materials should be investigated. 2) The wafer-scale growth of 2D materials needs to be highly developed than ever to make them more compatible with the silicon industry. 3) Stability, long-term toxicity, and reusability of 2D-based bio-FETs should be carefully studied in-depth, and 4) more attention would be paid to the performance of 2D-based bio-FETs in the clinical samples in order to be commercially competitive.

Despite these challenges, 2D materials have much more to offer as biosensors. 2D materials show high surface area, excellent electrical conductivity, long-term chemical stability and good biocompatibility. Their integration in FET biosensors offers low detection limits, flexibility, real-time monitoring, label-free diagnosis and high selectivity. The 2D based bio-FETs promise development of new point of care testing devices which are suitable for future telemedicine. In addition, 2D bio-FETs are usually associated with a top-down fabrication approach that is more compatible with current technology. This area is at the beginning of the road and is expected to make significant progress in the near future. Altogether, 2D bio-FET can be one of the greatest medical achievements and soon its prominent role in human daily life will be seen.

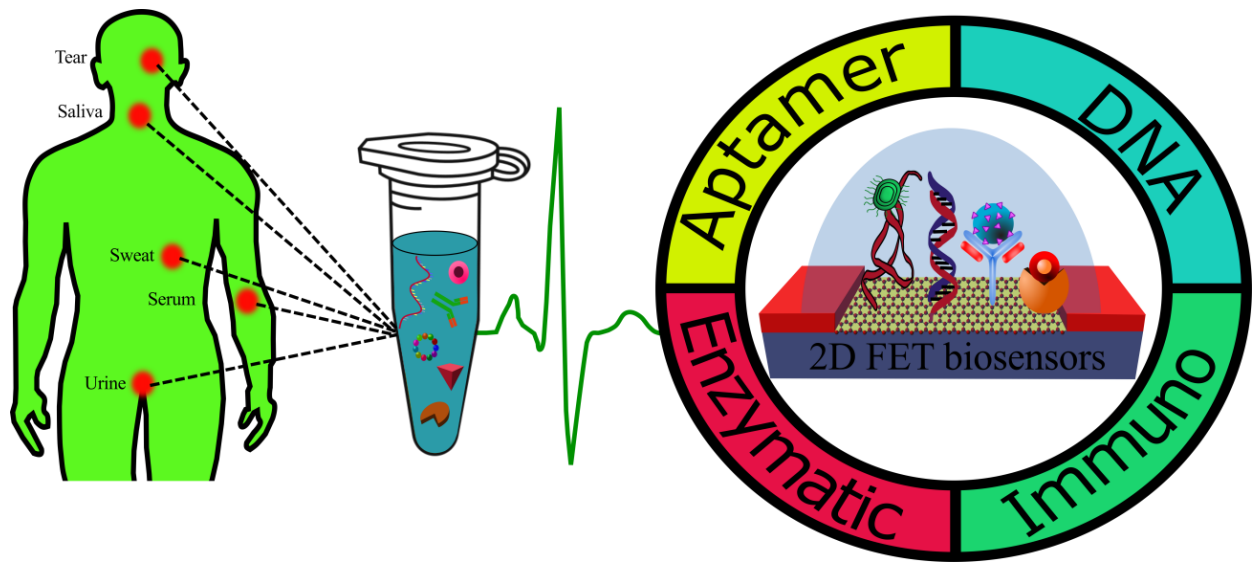

Graphical abstract

Highlights

- The field-effect transistor biosensors (bio-FETs) are presented based on the 2D materials.
- The principle of biosensing in the 2D based bio-FETs are discussed.
- Four categories of the 2D based bio-FETs are introduced depending on their bio-receptor components.
- The performance of the introduced bio-FETs is discussed in bio-medical samples.